\documentclass[%
 reprint,
 amsmath,amssymb,
 aps,
]{revtex4-2}

\usepackage{graphicx,epsfig,units}
\usepackage{xcolor} 
\usepackage{soul} 
\usepackage{chemformula}
\usepackage{multirow}
\usepackage{booktabs}
\usepackage[flushleft]{threeparttable}
\usepackage[super]{nth}
\usepackage{float} 
\usepackage{amsmath,amsfonts,mathrsfs,amsbsy,bm,babel}
\usepackage{natbib}
\usepackage[caption=false]{subfig}
\usepackage{dblfloatfix} 
\usepackage{array,booktabs,arydshln,xcolor}
\newcommand\VRule[1][\arrayrulewidth]{\vrule width #1}

\usepackage{hyperref}
\usepackage{appendix}
\makeatletter
\newcommand*{\rom}[1]{\expandafter\@slowromancap\romannumeral #1@}
\makeatother
\hypersetup{
    colorlinks=true,
    linkcolor=blue,
    filecolor=magenta,      
    urlcolor=black,
}

\begin{document}

\title{Exact Solution for Elastic Networks on Curved Surfaces}

\author{Yinan Dong}
\affiliation{Department of Physics and Astronomy, University of California Riverside, Riverside, California 92521, United States}

\author{Roya Zandi}
\affiliation{Department of Physics and Astronomy, University of California Riverside, Riverside, California 92521, United States}

\author{Alex~Travesset}
\affiliation{Department of Physics and Astronomy, Iowa State University and Ames Lab, Ames, Iowa 50011, United States}

\begin{abstract}

The problem of characterizing the structure of an elastic network constrained to lie on a frozen curved surface appears in many areas of science and has been addressed by many different approaches, most notably, extending linear elasticity or through effective defect interaction models. In this paper, we show that the problem can be solved by considering non-linear elasticity  in an exact form without resorting to any approximation in terms of geometric quantities. In this way, we are able to consider different effects that have been unwieldy or not viable to include in the past, such as a finite line tension, explicit dependence on the Poisson ratio or the determination of the particle positions for the entire lattice. Several geometries with rotational symmetry are solved explicitly. Comparison with linear elasticity reveals an agreement that extends beyond its strict range of applicability. Implications for the problem of the characterization of virus assembly are also discussed.

\end{abstract}
\pacs{}
\maketitle

Deciphering the design principles of life is one of the lingering mysteries facing researchers in many areas of science.  Among many biological systems, viruses, in particular, have received much more attention as they are ubiquitous pathogens in our environment with members infecting all kingdoms of life. Most viruses, from the simplest to the most complicated, and from the least to the most evolved, are constituted of a protein shell (or “capsid”) that encloses the viral genetic material (RNA or DNA) \cite{Hagan2013,Garmann2016}. Understanding the process of virus assembly is a fundamental challenge of ever-increasing interest, not only because it is a central stage of the viral life cycle, but also because it is the target of antiviral therapeutic strategies.  The Coronavirus Disease 2019 (COVID-19) pandemic, connected to SARS-CoV-2 revealed more than ever the importance of identifying new ways to combat viruses. In this context, our current understanding of virus assembly is quite limited. The difficulties arise from the interplay between curvature and crystalline order and their role in determining the positions of lattice defects on elastic surfaces with non-zero Gaussian curvature \cite{twarock2019structural,panahandeh2018equilibrium,panahandeh2020virus,mohajerani2018role}.

\begin{figure}[ht]
\includegraphics[width=0.81\linewidth]{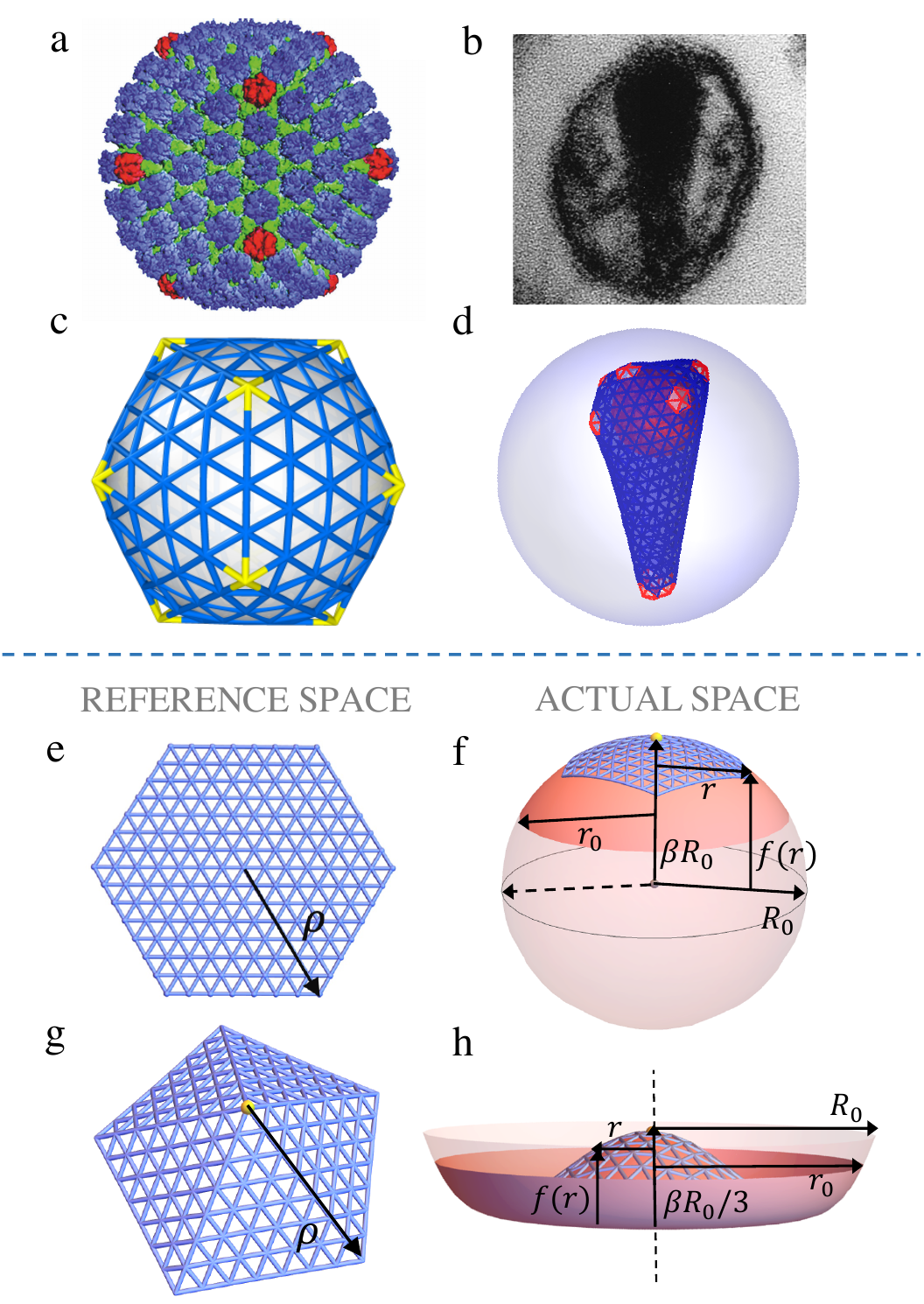}
\caption{(a) 3D cryo-EM reconstruction of HSV \cite{zhou2000seeing}; (b) EM image of HIV surrounded with the lipid envelope \cite{ganser1999assembly}. The result of the computer simulations of (c) an HSV \cite{LiTravesset2018} and (d) an HIV with genome and membrane \cite{nature2016}. (e) The reference space consists of a flat surface without a disclination or (g) with a disclination at the center. The actual space is any manifold endowed with its natural metric. As an example, we consider (f) a spheroid($\beta=1$) and (h) a sombrero($\beta=3$).}
\label{fig:fig1}
\end{figure}

Fig.~\ref{fig:fig1} illustrates the structure of two viruses with different geometries: (a) Herpes Simplex Virus (HSV) with icosahedral symmetry \cite{zhou2000seeing} and (b) Human Immunodeficiency Virus (HIV) \cite{ganser1999assembly} with a conical structure. In the case of HSV, the position of 12 pentagonal defects is precise (Fig.~\ref{fig:fig1}c) to preserve the symmetry of the shell. In previous work, we have shown that as a triangular lattice such as a HSV shell (Fig.~\ref{fig:fig1}c) \cite{LiTravesset2018}  grows over a spherical scaffold, defects appear one by one at the vertices of an icosahedron, explaining how error-free structures with icosahedral order assemble. For HIV shells, while there are often 5 defects at the smaller and 7 at the larger caps, defect positions can vary from one HIV structure to another. The computer simulations of Ref.~\cite{nature2016} have shown how the presence of genome and membrane contributes to the formation of the conspicuous conical HIV structures; nevertheless, currently, there is no clear understanding of what determines the position of the defects as the surfaces with non-zero Gaussian curvature such as the conical shell of HIV grow.

Capsid formation dynamics is just one example of the more general class of problems consisting of crystal growth on curved geometries. Other examples are faceted insect eyes, liquid crystals, curved array of microlenses in optical engineering systems, other protein cages in addition to viral capsids such as platonic hydrocarbons, heat shock proteins, ferritins, carboxysomes, silicages, multicomponent ligand assemblies, clathrin vesicles and many other cellular organelles, as reviewed in Ref.~\cite{Zandi2020}, for example. These problems have been mostly addressed by extending linear elasticity \cite{Seung1988, MorozovBruinsma2010,AzadiGrason2014,Grason2015,Castelnovo2017, Kosmrlj2017} or through effective defect interaction models \cite{Perez-GarridoDodgsonMoore1997,Bowick2000,BowickMe2002,Lidmar2003,GiomiBowick2007,Agarwal2020}. Either approach brings significant limitations: linear elasticity and its extensions impose certain approximations on geometric quantities such as Gaussian curvature (see discussion in SI Sec.~\rom{3}, specially Eq.~(S20)) and fail to satisfy global topological and geometrical constraints, most notably the Gauss Bonnet theorem \cite{LiTravesset2019}. Effective defect interaction models satisfy global properties exactly, but with uncontrolled approximations and obvious deficiencies: they predict a universal independence of all measurable quantities on the Poisson ratio $\nu_T$ and it is unknown, thus far, how to include other relevant free energy terms such as line tension. Furthermore, it requires computing the inverse square Laplacian, a formidable complex task beyond simple cases.

In this paper, we formulate general elasticity in curved surfaces based on geometric invariants and solve the equations exactly. We build upon our previous results for spherical caps~\cite{LiTravesset2019,Li2019} combined with covariant elasticity \cite{Efrati2009, MosheSharonKupferman2015}. Thus, the approach not only surmounts all the limitations of the previous theories, but provides an exact solution to the problem. Furthermore the theory allows to explore the impact of line tension and the Poisson ratio on the assembly of curved surfaces. The method is general and may be generalized to surfaces without specific symmetries. In this way, the approach provides the first step towards obtaining a complete theory for the self assembly of non-spherical virus capsids. 

The free energy of a partially formed elastic shell can be written as
\begin{eqnarray}\label{Eq:intro:free_dens}
   F &=& F^{elastic} + F^{bending} + F^{abs}+F^{line} \\ \nonumber 
   &=&\int d^2{\bm x}\sqrt{g}\left[{\cal F}^{elastic} + {\cal F}^{bending}\right]+F^{abs}+F^{line} \ .
\end{eqnarray}
where the first and second terms correspond to the stretching and bending energies of the patch respectively, the third term represents the attractive monomer-monomer interaction promoting the crystal growth and the last term is associated with the cost of the rim energy due to the fact that the subunits at the boundary have fewer neighbors than the ones at the interior of the surface. The elastic term ${\cal F}^{elastic}$ (see Eq.~\ref{Eq:app:elastic energy density}) contains a quadratic term \cite{LandauElasticityBook} in the strain tensor $u_{\alpha\beta}$ 
\begin{equation}\label{Eq:intro:strain}
    u_{\alpha\beta}=\frac{1}{2}\left[g_{\alpha \beta}-\bar{g}_{\alpha \beta}\right] \ ,
\end{equation}
where $g_{\alpha\beta}({\bm x})$ is the {\em actual metric} (the metric of the curved surface) and $\bar{g}_{\alpha\beta}(\bar{\bm x})$ is the {\em reference metric} describing a perfect lattice, {\it i.e.} the one consisting of equilateral triangles (Fig.~\ref{fig:fig1}e-h). Note that the Gaussian curvature of the reference metric is zero, except possibly, on a finite number of points that define disclination cores \cite{LiTravesset2019}. If $a_L$ is the lattice constant of the perfect triangular lattice in reference space, the number of particles $N >> 1$ making the crystal with a given area $\hat{A}$ is 
\begin{equation}\label{Eq:intro:area_reference}
    \hat{A}=\int d^2{\bm x}\sqrt{\bar{g}({\bm x})}=\int d^2{\bar{\bm x}}\sqrt{\bar{g}(\bar{\bm x})}=\frac{\sqrt{3}}{2} N a_L^2 \ .
\end{equation}
The second term in Eq.~\ref{Eq:intro:free_dens} can be expressed in terms of the two radii of curvature $(R_i)_{i=1,2}$,
\begin{equation}\label{Eq:intro:free_bending}
   {\cal F}^{bending} = \kappa\left[ \left(\frac{1}{R_1}-H_0\right)^2+ \left(\frac{1}{R_2}-H_0\right)^2\right]  \ ,
\end{equation}
with $\kappa$ the bending rigidity and $H_0$ the mean spontaneous curvature of the constituents or subunits. We emphasize that the free energy density, Eq.~\ref{Eq:intro:free_dens}, has no trivial solution. The only surfaces allowing zero strains have either zero Gaussian curvature: a plane or cylinder ($q=0$) or a delta function at the origin like a cone ($q=1$). The absolute minimum of the bending rigidity implies a surface with $R_1=R_2=\frac{1}{H_0}$, a sphere. There is no surface that simultaneously minimizes both the elastic and bending energies. The third term in Eq.~\ref{Eq:intro:free_dens}, $F^{abs}=-\Pi \hat{A} < 0$ with $\Pi$ the attractive interaction per unit area due to favorable hydrophobic contacts between subunits, is the driving force for crystal growth \cite{MorozovBruinsma2010}. 

In this paper we consider a frozen geometry, so the 
actual metric $g_{\alpha\beta}({\bm x})$ is that of the corresponding surface. The defect distribution is also given, so the reference metric $\bar{g}_{\alpha\beta}(\bar{\bm x})$ is fixed as well. The problem consists of mapping the actual and reference space ${\bm x}={\cal U}(\bar{{\bm x}})$. In other words, ${\cal U}$ is the function that determines how the perfect lattice in reference space maps to the deformed one in actual space, as illustrated in Fig.~\ref{fig:fig1}e-h. For simplicity, in this paper we consider only surfaces of revolution, defined by $x=r \cos(\theta), y=r \sin(\theta),z=f(r)$ with actual metric
\begin{equation}\label{Eq:intro:actual_metric}
    ds^2=\left(1+f^{\prime}(r)^2\right)dr^2+ r^2 d\theta^2 \ ,
\end{equation}
see Fig.~\ref{fig:fig1}.  Further, we consider an isotropic reference metric
\begin{equation}\label{Eq:intro:reference_metric}
ds^2 = \rho^{\prime}(r)^2 dr^2+ \alpha^2 \rho(r)^2 d\theta^2 = d^2\rho+\alpha^2\rho^2 d\theta^2
\end{equation}
with $\alpha=1-\frac{q}{6}$ and $q=0, \pm 1$ corresponding to no disclination or with a disclination at the origin. The two metrics are basically {\em incompatible} \cite{MosheSharonKupferman2015}, that is, for a fixed $f(r)$ there is no choice of $\rho(r)$ that will make the strain tensor Eq.~\ref{Eq:intro:strain} vanish, as their Gaussian curvatures generally differ.  In the isotropic case,  the function ${\cal U}$ describing the mapping from the actual to reference space (or reference to actual) can be expressed as $\rho(r) \mbox{ or }  r(\rho)$.
To make the presentation of the paper simple, we consider a situation in which the surface is given through $f(r)$ (see Fig.~\ref{fig:fig1}). The elastic energy given in Eq.~\ref{Eq:intro:free_dens} denpends on $\rho(r)$ and thus becomes Eq.~\ref{Eq:app:elastic energy}. And then the problem consists of finding the optimal $\rho(r)$ 
that minimizes the free energy Eq.~\ref{Eq:intro:free_dens}. Following Ref.~\cite{LiTravesset2019,Efrati2009}, this leads to
\begin{equation}\label{Eq:intro:GM:covariant}
    \nabla_{\alpha} \sigma^{\alpha \beta}+(\bar{\Gamma}^{\beta}_{\gamma \nu}-{\Gamma}^{\beta}_{\gamma \nu})\sigma^{\gamma \nu} = 0 \ ,
\end{equation}
where $\sigma^{\alpha\beta}$ is the stress tensor and ${\Gamma}^{\beta}_{\gamma \nu}$ are the Christoffel symbols for the reference and actual metrics (see Appendix~\ref{app:B}). This is a one dimensional non-linear differential equation with just one unknown $\rho(r)$ (see Eq.~\ref{app:explicit diff}).  We solve Eq.~\ref{Eq:intro:GM:covariant} subject to the following boundary condition, 
\begin{equation}\label{Eq:intro:boundary}
    n_{\rho}\sigma^{\rho \lambda}\bar{g}_{\lambda \nu} = -\frac{\tau}{r_A} n_{\nu} \ .
\end{equation}
where $\tau$ is the line tension, $n^{\mu}=g^{\mu \nu} n_{\nu}$ is the normal to the boundary within the surface and $r_A$ is the curvature of the boundary. For a surface with rotational symmetry and a circular boundary $r = r_0$, this equation simply becomes Eq.~\ref{Eq:app:boundary condition with rotational symmetry}. For a tensionless boundary, obviously $\tau=0$. In Appendix~\ref{app:C}, we provide a detailed derivation of Eq.~\ref{Eq:intro:boundary} from the line energy $F^{line}= \tau \oint_{\partial D} ds$ with $ds$ an infinitesimal length for the boundary ${\partial D}$ in actual space. We also show how Eqs.~\ref{Eq:intro:free_dens}-\ref{Eq:intro:boundary} reduce to standard linear elasticity and provide an explicit analytical solutions within linear elasticity for both a spheroid, $f(r)=\beta \sqrt{R_0^2-r^2}$ (Fig.~\ref{fig:fig1}f) and a sombrero surface, $f(r)={\beta R_0}/3\left(1-\left({r}/{R_0}\right)^2+\left({r}/{R_0}\right)^4\right)^{3/2}$ (Fig.~\ref{fig:fig1}h), where $\beta$ is a unitless number, in SI Sec.~\rom{3}.

\begin{figure}[ht]
  \includegraphics[width=.98\linewidth]{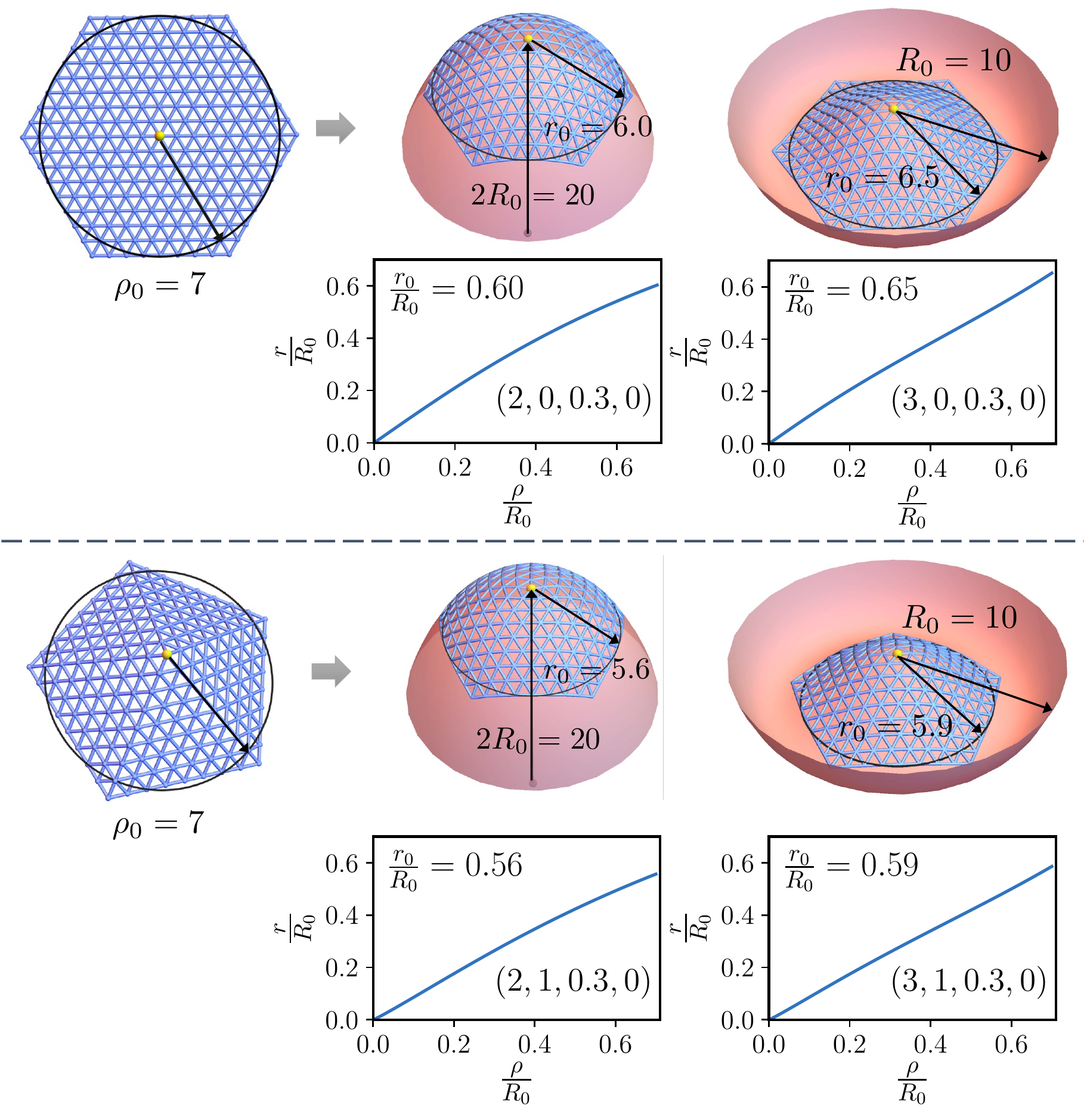}
\caption{Lattice reconstruction for both the spheroid and sombrero. The lengths $R_0$, $\rho_0$ and $r_0$ are given in units of the lattice constant $a_L$. The plots under spheroid and sombrero surfaces indicate the corresponding mapping function from the reference to actual space. $\frac{\rho_0}{R_0}=0.7$ for each plot. The number in the parenthesis in each figure denote $(\beta, q_i, \nu_p, \hat{\tau})$. }
\label{fig:lattic-recon}
\end{figure}

To obtain the free energy of the system, we first calculate $\rho(r)$ through Eq.~\ref{Eq:intro:GM:covariant}, which minimizes Eq.~\ref{Eq:intro:free_dens}. The plots in Fig.~\ref{fig:lattic-recon} below each show the optimal $\rho(r)$ or $r(\rho)$ for both spheroid and sombrero surfaces with no disclination or with one disclination at the center.  It is important to note that with the exact $r(\rho)$, we can reconstruct the lattice in actual space; the positions of the lattice in reference space $(\rho_i,\theta_i)_{i=1\cdots N}$ are known and consists of $N$ equilateral triangles with lattice constant $a_L$ (if $N>>1$), see Eq.~\ref{Eq:intro:area_reference}. Then, from $r(\rho)$, the positions in actual space $(r(\rho_i),\theta_i)_{i=1\cdots N}$ are obtained, as illustrated in Fig.~\ref{fig:lattic-recon}. Thus, we find a solution to the problem of finding the best possible triangulation consisting of equilateral triangles that cover a given surface. Note that this solution is independent of the underlying potential among the constituent particles, and therefore, hereon we refer to this triangulation as the {\em universal mapping lattice}.

\begin{figure}[ht]
\includegraphics[width=.98\linewidth]{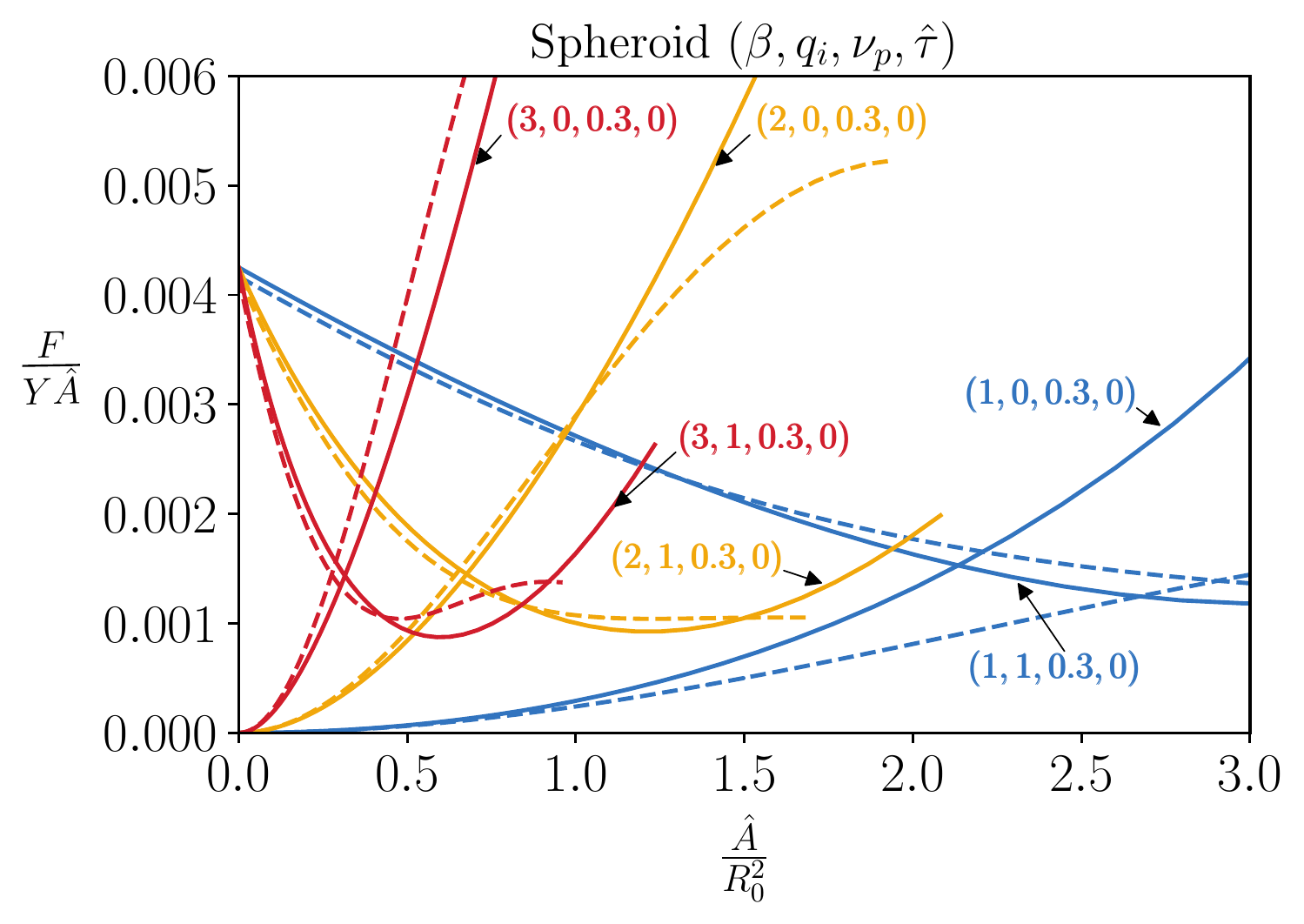}
\includegraphics[width=.98\linewidth]{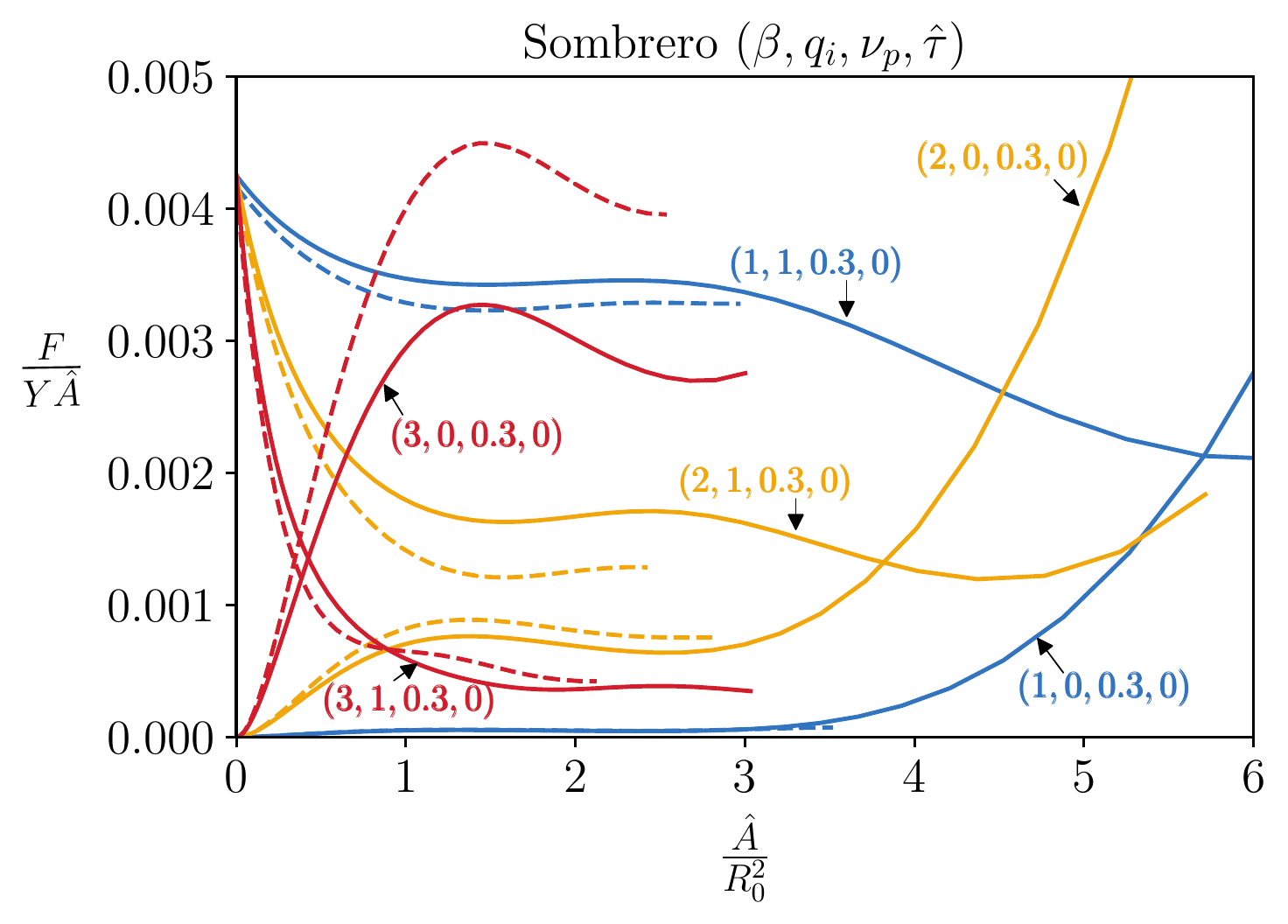}
\caption{Free energy for a spheroid and sombrero without ($q_i=0$) and with ($q_i=1$) a disclination at the center at  $\frac{\tau}{Y R_0}\equiv \hat{\tau}=0$ (zero line tension) and at fixed Poisson ratio $\nu_T=0.3$. The solid line corresponds to the exact results while the dashed line denotes the analytical results within linear elasticity. The three different colors represent different values of $\beta=1, 2, 3$, indicating the magnitude of f(r).}
\label{fig:fig2}
\end{figure}

Plugging the solutions of $\rho(r)$ into Eq.~\ref{Eq:intro:free_dens}, we obtain the free energy of the system. First, we consider the case of free boundary conditions as illustrated in Fig.~\ref{fig:fig2}. For comparison, we show the predictions from linear elasticity, which become exact in the limit of small curvature ($\frac{\hat{A}}{R^2_0} \rightarrow 0$), both for the defect free case $q_i=0$ and a single disclination $q_i=1$. Very generally, we find that the applicability of elasticity theory extends to relatively large curvatures ($\frac{\hat{A}}{R^2_0} \approx 1$). For the spheroid, linear elasticity remains qualitatively correct for the entire range explored, but this is not the case for the sombrero surface, see Fig.~\ref{fig:fig2} , where linear elasticity breaks down and cannot be extended beyond a certain limit.

\begin{table}[]
    \centering
    
    \begin{tabular}{!{\VRule}c!{\VRule[1pt]}c!{\VRule}c!{\VRule}c!{\VRule[1pt]}c!{\VRule}c!{\VRule}c!{\VRule[1pt]}c!{\VRule}c!{\VRule}c!{\VRule}}
         \specialrule{0.5pt}{0pt}{0pt}
         & \multicolumn{3}{c!{\VRule[1pt]}}{$\beta=1$} & \multicolumn{3}{c!{\VRule[1pt]}}{$\beta=2$} & \multicolumn{3}{c!{\VRule}}{$\beta=3$}  \\
         \specialrule{0.5pt}{0pt}{0pt}
         & $\frac{\hat{A}}{R^2_0}$ & $\theta$ & $\theta_{AH}$ & $\frac{\hat{A}}{R^2_0}$ & $\theta$ & $\theta_{AH}$ & $\frac{\hat{A}}{R^2_0}$ & $\theta$ & $\theta_{AH}$ \\
        \specialrule{0.5pt}{0pt}{0pt}
      Spheroid  &  2.13 &  0.75 & 0.73 &  0.68 &  0.43 & 0.36 &  0.32 &  0.29 & 0.27 \\
      \specialrule{0.5pt}{0pt}{0pt}
        Sombrero & 5.71 &  1.18 & 1.17 &  3.83 &  1.00 & 0.89 &  0.42 &  0.34 & 0.35 \\
        \specialrule{0.5pt}{0pt}{0pt}
    \end{tabular}

    \caption{Transition points at which the surface with ($q_i=1$) and without ($q_i=0$) a disclination have the same energy. Here $\theta=\frac{r}{R_0}$ and $\theta_{AH}$ are the predictions reported in Ref.~\cite{Agarwal2020}. }
    \label{tab:disc_transition}
\end{table}

The points where the $q_i=0$ and $q_i=1$ curves cross each other define whether it is energetically favorable to have a disclination at the center or not. The results are 
quoted in Table~\ref{tab:disc_transition}. We compare with the most recent, and to our knowledge, most accurate predictions from Ref.~\cite{Agarwal2020}. Note that in some cases the differences are as high as 10\%.

\begin{figure}[ht]
  \includegraphics[width=.98\linewidth]{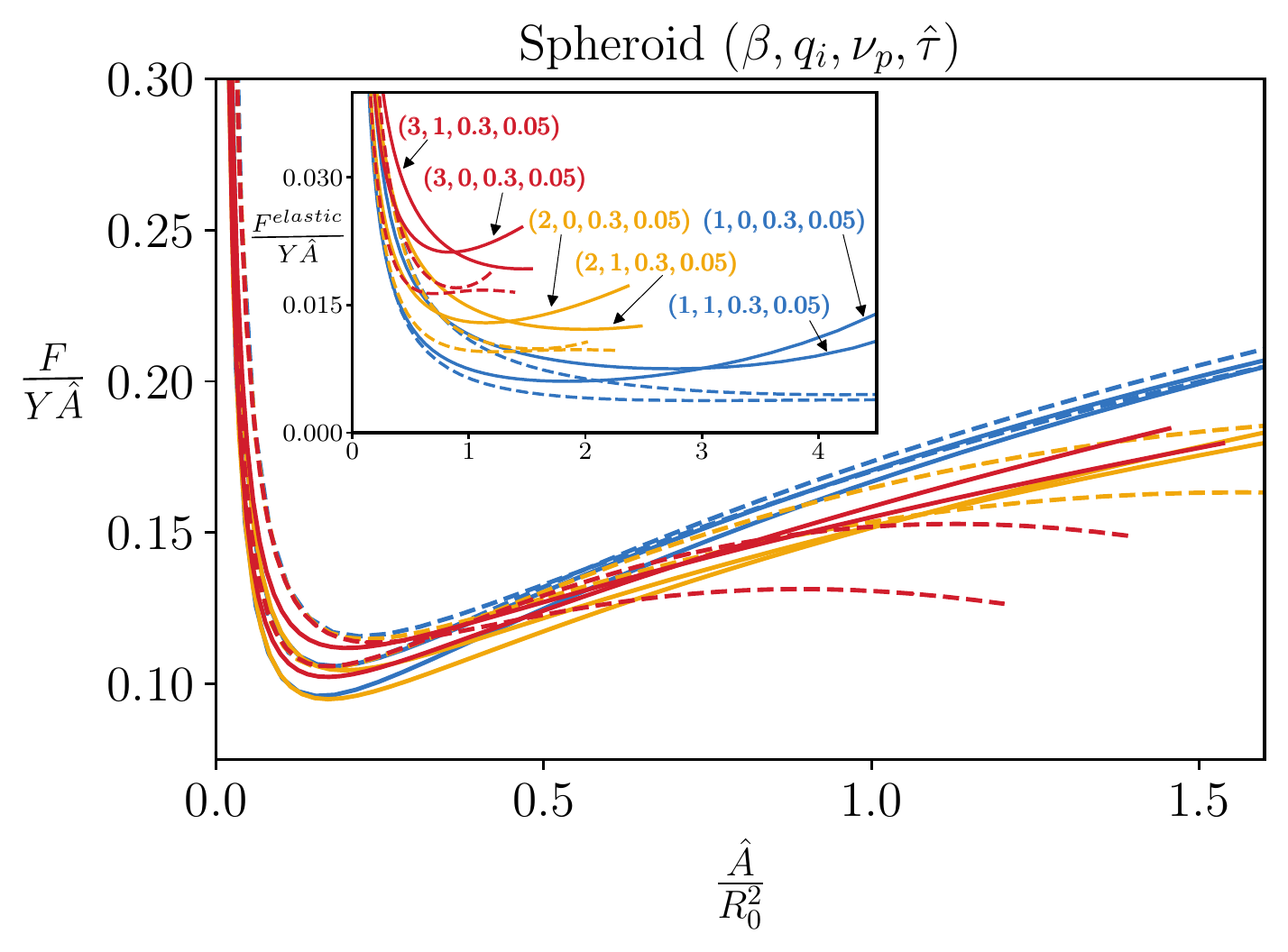}
  \includegraphics[width=.98\linewidth]{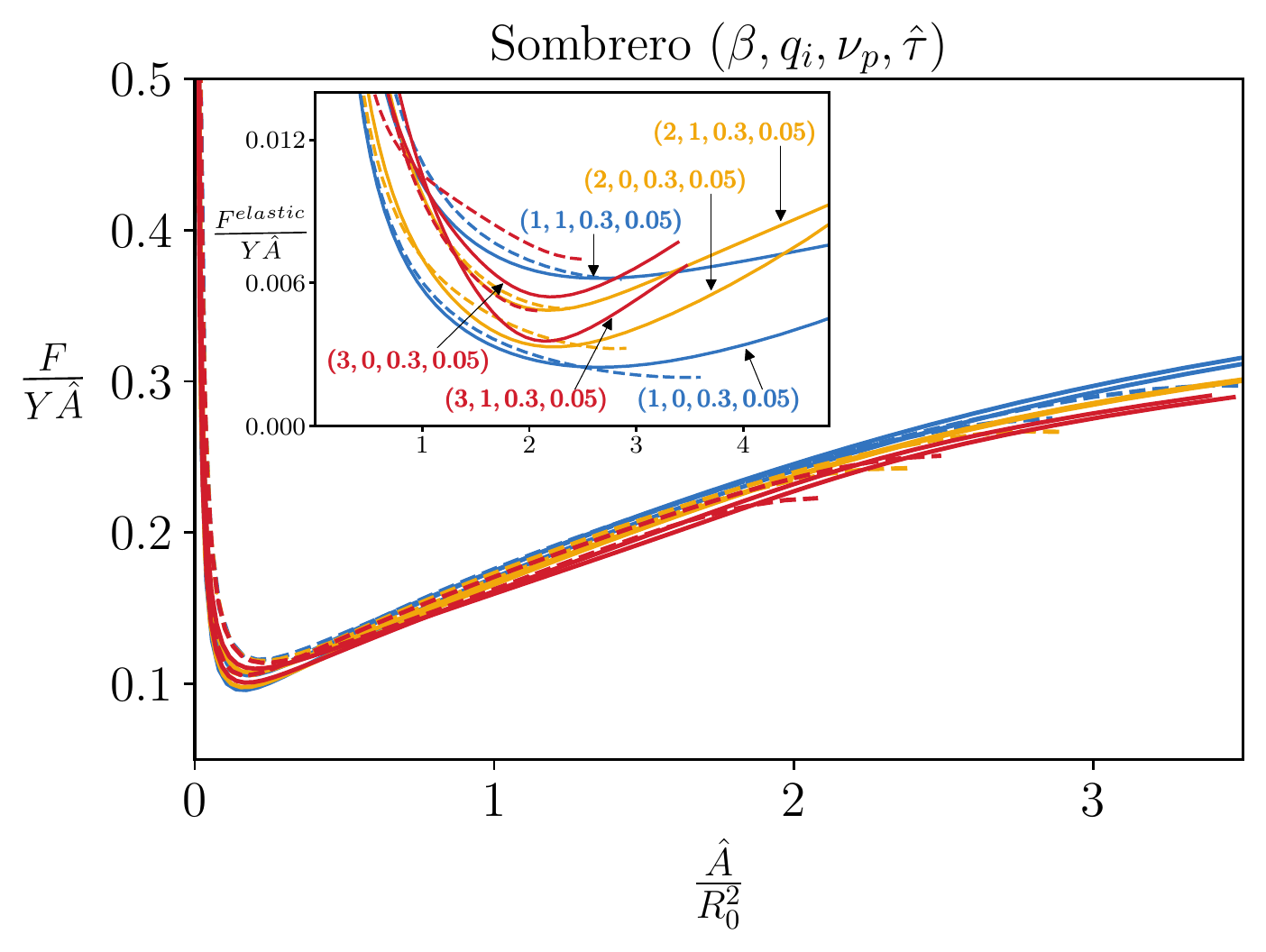}
\caption{Free energy (Eq.~\ref{Eq:intro:free_dens}) for both the spheroid and sombrero at finite line tension $\frac{\tau}{Y R_0}=\hat{\tau}=0.05$. The solid line corresponds to the exact results while the dashed line denotes the analytical results within linear elasticity.}
\label{fig:fig3}
\end{figure}

The case of a finite line tension is shown in Fig.~\ref{fig:fig3}. Basically, a large value of $\tau$ overshadows all the other energies of the system resulting into the collapse of all the free energy plots into an almost universal curve defined by the line tension term. The overall free energy is in some agreement with linear elasticity theory, which, as shown in the inset, is also true for the elastic contribution, see Eq.~\ref{Eq:intro:free_dens}. In SI Sec.~\rom{5}, we provide similar plots for smaller values of the line tension (see Fig.~S1).

\begin{figure}[ht]
  \includegraphics[width=.98\linewidth]{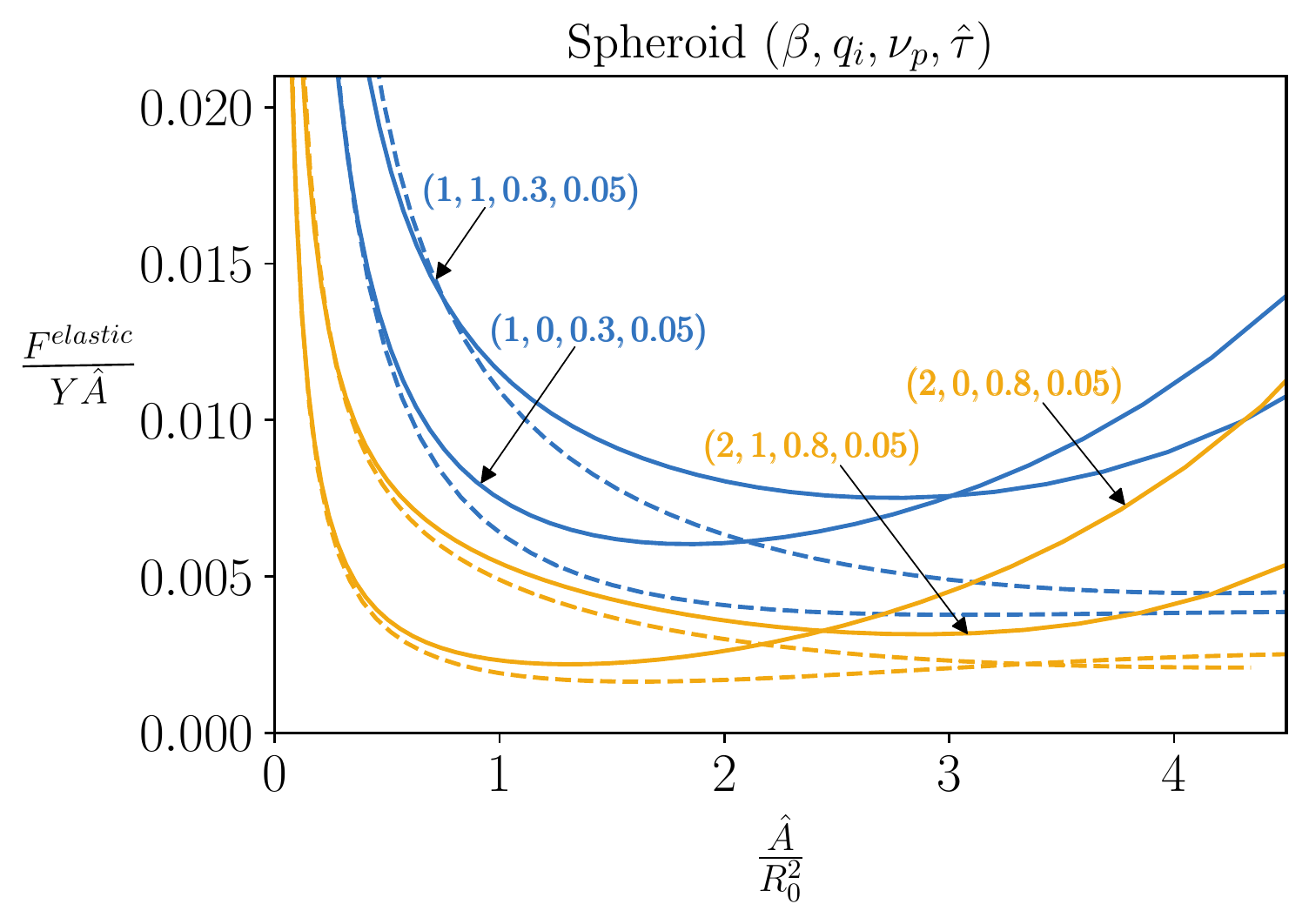}
\caption{The elastic energy (see Eq.~\ref{Eq:intro:free_dens}) for the spheroid at $\frac{\tau}{Y R_0}=\hat{\tau}=0.05$ for two different values of the Poisson ratio $\nu_T=0.3$ and $0.8$. The solid lines correspond to the exact results while the dashed lines denote analytical results within linear elasticity.}
\label{fig:fig4}
\end{figure}

In the absence of line tension, linear elasticity predicts that the free energy is a function of the Young modulus $Y$ alone, independent of the Poisson ratio $\nu_T$ in SI Sec.~\rom{3}. This is implicitly assumed in models of interacting defects. Within the exact non-linear theory presented in this paper, we show that the dependence of the Poisson ratio at vanishing line tension is, indeed, negligible (see Fig.~S2). However, Fig.~\ref{fig:fig4} reveals that there is a dependence of the Poisson ratio whenever the line tension is non-zero, see also Fig.~S2.  We have been able to extract explicitly the dependence of $\rho(r)$ or $r(\rho)$ on the Poisson ratio as the solution of the exact theory, see Figs.~\ref{fig:lattic-recon},~S3 and ~S4.

As expected, for $\frac{\hat{A}}{R_0^2}$ small, the results are well described by linear elasticity, but as this value increases, they get progressively worse 
, and in the case of the sombrero, linear elasticity breaks down for sufficiently large 
values of $\frac{\hat{A}}{R_0^2}$.

In summary, this paper provides an exact solution to the problem of determining the structure of crystals on a curved surface by formulating the problem in terms of geometric invariants, which besides connecting it to the well developed field of differential geometry of curves and surfaces, enables, what we believe, is a transparent interpretation of non-linear elasticity theory. In addition, we show that effects that have been difficult to consider in the past, such as a finite line tension or dependence on the Poisson ration $\nu_T$ are easily included.

Our exact solution provides a {\em universal triangulation}, as shown in Fig.~\ref{fig:lattic-recon}, that is, a solution to the problem of providing the optimal tiling of an arbitrary surface with triangles as close to equilateral as possible. In previous studies, determining particle positions requires numerical minimization methods, either through discretizations of elasticity theory \cite{Seung1988,Travesset2003,Li2019} or using explicit potentials, most typically inverse power laws, on a sphere \cite{PerezGarrido1996,BowickMe2002} and other geometries \cite{BenditoBowickMedinaYao2013} as well. The problem with numerical minimizations is the cost and instabilities that appear for both a large number of particles $N$ and/or complicated geometries. We note that our analytical approach does not suffer from any of these problems: it is independent of the number of particles $N$, see Eq.~\ref{Eq:intro:area_reference}, and is stable for any differentiable surface.

The approach developed in this paper allows us to understand how the pentameric defects appear and interact with each other with clear implications in viral shells, specifically for the difficult cases of the assembly of nonspherical structures similar to those presented in Fig.~\ref{fig:fig1}. Determining the location of lattice defects in the growing shells with non-zero Gaussian curvature in the presence of boundaries with line tension for various values of Poisson ratio has been shown to be a very challenging task \cite{Garmann201909223,Chevreuil2018,dykeman2014solving,Panahandeh2022}. The theory developed here paves the path for tackling the problem of crystalline growth pathways. While in this paper we restricted our study to a fixed geometry with a given number of particles and provided explicit solutions for problems with rotational symmetry, the approach is completely general for any geometry or number of disclinations, although its explicit description requires additional developments that will be provided in subsequent studies. 

\begin{acknowledgments}

The work of AT is funded by NSF, DMR-1606336. RZ and YD acknowledge support from NSF DMR-2131963 and the University of California Multicampus Research Programs and Initiatives (grant No. M21PR3267).
\end{acknowledgments}

\begin{appendix}

\section{\label{app:A}Explicit Formulas for the Different Quantities}
In this appendix, we provide explicit expressions for the formulas in the main text.

The explicit form of ${\cal F}^{elastic}$ is
\begin{equation}\label{Eq:app:elastic energy density}
    {\cal F}^{elastic}=\frac{1}{2} A^{\alpha\beta\gamma\delta} u_{\alpha \beta} u_{\gamma \delta},
\end{equation}
where $u_{\alpha\beta}$ (see Eq.~\ref{Eq:intro:strain} in the main text) is the strain tensor and 
\begin{equation}\label{Eq:app:EF:Elastic_constant}
    A^{\alpha\beta\gamma\delta}=\frac{Y}{1-\nu^2_p}\left[\nu_p g^{\alpha\beta}g^{\gamma\delta}+(1-\nu_p)g^{\alpha\gamma}g^{\beta\delta} \right] \ ,
\end{equation}
with $Y$ the Young Modulus, $\nu_p$ the Poisson ratio and $g^{\alpha\beta}$ the actual metric.

The Gaussian curvature is
\begin{equation}\label{Eq:app:EF:gaussian}
    K = \frac{\mbox{det}(\partial_i\partial_j f)}{(1+(\nabla f)^2}=\frac{f^{\prime}(r)f^{\prime\prime}(r)}{r(1+f^{\prime}(r)^2)^2} \ ,
\end{equation}
and the mean curvature (with the convention that $R_i=R>0,i=1,2$ for the sphere) is
\begin{eqnarray}\label{Eq:app:EF:Mean}
    2H &=& -\nabla \cdot \left( \frac{\nabla f}{(1+(\nabla f)^2)^{1/2}} \right)\\ \nonumber
    &=&-\left( \frac{f^{\prime\prime}(r)}{(1+f^{\prime}(r)^2)^{3/2}}+\frac{f^{\prime}(r)}{r(1+f^{\prime}(r)^2)^{1/2}} \right)
\end{eqnarray}
see Ref.~\cite{Zandi2020} for the details. The two curvatures can be obtained from the equation
\begin{eqnarray}
    K&=&\frac{1}{R_1 R_2} \quad \nonumber\\
    2H&=&\frac{1}{R_1}+\frac{1}{R_2} \ ,
\end{eqnarray}
such that $\frac{1}{R_1}=H+\sqrt{H^2-K}$ and $\frac{1}{R_2}=H-\sqrt{H^2-K}$ with $H$ and $K$ given in Eqs.~\ref{Eq:app:EF:gaussian} and ~\ref{Eq:app:EF:Mean}.
The stress tensor is
\begin{equation}\label{Eq:app:EF:Stress}
    \sigma^{\alpha\beta}=\frac{1}{\sqrt{g}}\frac{\delta F}{\delta u_{\alpha \beta}}=A^{\alpha\beta\gamma\delta}  u_{\gamma \delta} \ .
\end{equation}

\section{\label{app:B}Particularization to Surface Revolution}

 In this section, we provide various quantities for the surfaces of revolution, defined by $x=r \cos(\theta), y=r \sin(\theta),z=f(r)$ with actual metric defined in Eq.~\ref{Eq:intro:actual_metric}.

The nonzero Christoffel symbols are given in Ref.~\cite{LiTravesset2019} and are
\begin{eqnarray}\label{Eq:app:EF:Christoff:revolution}
    & &\mbox{symbol}\qquad\qquad\quad {\Gamma}^{r}_{rr}\qquad \qquad\quad {\Gamma}^{r}_{\theta\theta}\qquad\qquad\quad {\Gamma}^{\theta}_{\theta r} \nonumber \\
    & &\mbox{reference} \qquad\quad\;\; \frac{\rho''(r)}{\rho'(r)} \qquad\quad -{\alpha}^{2}\frac{\rho(r)}{\rho'(r)}\quad\qquad \frac{\rho'(r)}{\rho(r)} \nonumber \\
    & &\mbox{actual} \qquad\quad\;\;\; \frac{f'(r)f''(r)}{1+f'(r)^2}\qquad\quad\!\!\!\!\!\frac{-r}{1+f'(r)^2}\qquad\quad \frac{1}{r}\nonumber
\end{eqnarray}

The elastic energy given in Eq.~\ref{Eq:intro:free_dens} depends on $\rho(r)$, 
\begin{eqnarray}\label{Eq:app:elastic energy}
    &&\frac{F^{elastic}}{Y\pi} = \frac{1}{4(1-\nu^2_p)}\int dr r(1+f^{\prime}(r)^2)^{1/2}\\ \nonumber
    &&\left[ \left( 1-\frac{\rho^{\prime}(r)^2}{1+f^{\prime}(r)^2}\right)^2 
    +\left(1-\frac{\alpha^2 \rho(r)^2}{r^2}\right)^2 +\right. \\ \nonumber
    &&+ \left.2\nu_p\left( 1-\frac{\rho^{\prime}(r)^2}{1+f^{\prime}(r)^2}\right)\left(1-\frac{\alpha^2 \rho(r)^2}{r^2}\right) \right] 
\end{eqnarray}

The stress tensor given in Eq.~\ref{Eq:app:EF:Stress} becomes,
\begin{eqnarray}\label{Eq:app:EF:stress:revolution}
\sigma^{rr} &=& \frac{Y}{2(1-\nu^2_{p})(1+f'(r)^2)}\left[1-\frac{\rho'(r)^2}{1+f'(r)^2}+ \right. \\ \nonumber 
 &+&\left.\nu_{p}\left(1-\left(\frac{\alpha \rho(r)}{r}\right)^2\right)\right]  \\ \nonumber
\sigma^{r\theta} &=& \sigma^{\theta r} = 0 \\ \nonumber
\sigma^{\theta\theta} &=& \frac{Y}{2r^2(1-\nu^2_{p})}\left[\nu_{p}\left(1-\frac{\rho'(r)^2}{1+f'(r)^2}\right) + \right.\\ \nonumber
&+& \left. 1-\left(\frac{\alpha \rho(r)}{r}\right)^2\right] 
\end{eqnarray}

Using Eq.~\ref{Eq:app:EF:Christoff:revolution} and Eq.~\ref{Eq:app:EF:stress:revolution}, the general form of Eq.~\ref{Eq:intro:GM:covariant} for $\beta=r$ becomes
\begin{eqnarray}
    \partial_r\sigma^{rr}+\bar\Gamma^r_{rr}\sigma^{rr}+\bar\Gamma^r_{rr}\sigma^{rr}+\bar\Gamma^r_{\theta \theta}\sigma^{\theta \theta} +  \\ \nonumber
    + \Gamma^r_{rr}\sigma^{rr} + \Gamma^{\theta}_{\theta r}\sigma^{rr}  = 0,
\end{eqnarray}
which can be written as
\begin{eqnarray}\label{Eq:app:metric:diff}
   \frac{d \sigma^{rr}}{dr} + \left(\bar\Gamma^r_{rr}+\Gamma^r_{rr}+\Gamma^\theta_{\theta r}\right)\sigma^{rr}+\bar\Gamma^r_{\theta\theta}\sigma^{\theta\theta}=0 
\end{eqnarray}

The explicit form of the derivative of the stress tensor is,
\begin{eqnarray}
    \frac{d \sigma^{rr}}{d r}&=&\frac{-Y}{(1-v^2_p)}\frac{f'(r)f''(r)}{(1+f'(r)^2)^2}\left[1-\frac{\rho'(r)^2}{1+f'(r)^2} + \right. \\ \nonumber &+& \left. \nu_p\left(1-\left(\frac{\alpha \rho(r)}{r}\right)^2\right)\right]+\\\nonumber
    &+&\frac{Y}{(1-v^2_p)(1+f'(r)^2)}\left[\frac{\rho'(r)^2f'(r)f''(r)}{(1+f'(r)^2)^2}-\right. \\ \nonumber
    &-& \left. \frac{\rho'(r)\rho''(r)}{1+f'(r)^2}+\nu_p\frac{\alpha^2 \rho(r)^2}{r^3}-\nu_p\frac{\alpha^2\rho(r)\rho'(r)}{r^2}\right]
\end{eqnarray}

The equation determining $\rho(r)$, Eq.~\ref{Eq:app:metric:diff},  becomes
\begin{eqnarray}\label{app:explicit diff}
    &&\frac{-f'(r)f''(r)}{(1+f'(r)^2)}\left[1-\frac{\rho'(r)^2}{1+f'(r)^2}+\right.  \\ \nonumber
    &+&\left.\nu_p\left(1-\left(\frac{\alpha \rho(r)}{r}\right)^2\right)\right]+\frac{1}{1+f'(r)^2}\times \\ \nonumber
    &\times& \left[\frac{\rho'(r)^2f'(r)f''(r)}{(1+f'(r)^2)^2}-\frac{\rho'(r)\rho''(r)}{1+f'(r)^2}+\nu_p\frac{\alpha^2 \rho(r)^2}{r^3}-\right.\\ \nonumber
    &-&\left.\nu_p\frac{\alpha^2 \rho(r)\rho'(r)}{r^2}\right]+
    \left(\frac{\rho''(r)}{\rho'(r)}+\frac{f'(r)f''(r)}{1+f'(r)^2}+\frac{1}{r}\right)\times \\ \nonumber
    &\times&\frac{1}{2(1+f'(r)^2)}\left[1-
    \frac{\rho'(r)^2}{1+f'(r)^2}+\right.\\\nonumber
    &+&\left.\nu_p\left(1-\left(\frac{\alpha \rho(r)}{r}\right)^2\right)\right]-\frac{\alpha^2 \rho(r)}{2r^2 \rho'(r)}\times\\\nonumber
    &\times&\left[\nu_p\left(1-\frac{\rho'(r)^2}{1+f'(r)^2}\right)+1-\left(\frac{\alpha \rho(r)}{r}\right)^2\right] = 0\ .
\end{eqnarray}

\section{\label{app:C}Boundary Condition}
The boundary conditions can be obtained through the variations of $F^{area} = F^{elastic} + F^{bending}$ in Eq.~\ref{Eq:intro:free_dens} and the reparameterizations of the actual metric,
\begin{equation}\label{Eq:variation:reparam_ractual}
    \delta F^{area} = -\int d^2{\bm x}\partial_{\rho}\left(\sqrt{g}\sigma^{\rho \mu}\bar{\xi}_{\mu}\right) =\oint \sqrt{g} dx^{\mu}\varepsilon_{\mu\nu} \sigma^{\nu \rho} \bar{\xi}_{\rho} \ .
\end{equation}

If there is a line tension $\tau$, the contribution to the free energy is then
\begin{equation}\label{Eq:app:line energy}
    F^{line} =\tau \oint ds = \tau \oint \sqrt{\langle g \rangle} dl=\tau \oint dx^{\mu} g_{\mu\nu} t^{\nu} \ ,
\end{equation}
where $\langle g \rangle=g_{\mu\nu} \frac{d x^{\mu}}{dl} \frac{d x^{\nu}}{dl}$ and
\begin{equation}
    t^{\mu}=\frac{1}{\sqrt{\langle g \rangle}} \frac{dx^{\mu}}{dl},
\end{equation}
is the unit vector (in the actual metric) tangent to the boundary curve with $dx^{\mu}=\sqrt{\langle g \rangle} t^{\mu} dl$. The variations to the free energy Eq.~\ref{Eq:app:line energy} as shown in Ref.~\cite{LiTravesset2019} then become,
\begin{eqnarray}
    \delta F^{line}&=&\frac{\tau}{2} \oint dx^{\mu} \delta g_{\mu \nu} t^{\nu} =- \tau  \oint dx^{\mu} \nabla_{\mu}\xi_{\nu} t^{\nu}\\\nonumber
    &=&\tau  \oint dx^{\mu} \xi_{\nu} \nabla_{\mu} t^{\nu} 
\end{eqnarray}
Since $\delta\left(F^{area}+F^{line}\right)=0$, the appropriate boundary condition is then,
\begin{equation}
    \tau g_{\rho \nu} \nabla_{\mu} t^{\rho} = -\sqrt{g} \epsilon_{\mu \rho} \sigma^{\rho \lambda} \bar{g}_{\lambda \nu} \ .
\end{equation}
Note the tangent vector is 
\begin{equation} \label{Eq:tangentvector}
    t^{\rho}\nabla_{\rho} t^{\mu} =-\frac{1}{r_A} n^{\mu} \ ,
\end{equation}
where $r_A$ is the curvature of the curve defining the boundary. The normal to the tangent vector can be written as
\begin{equation}
     n_{\mu} = \sqrt{g} \varepsilon_{\mu \rho} t^{\rho}. 
\end{equation}
The boundary condition, thus, becomes
\begin{equation}
    n_{\rho}\sigma^{\rho \lambda}\bar{g}_{\lambda \nu} = -\frac{\tau}{r_A} n_{\nu} \ .
\end{equation}
This condition is slightly different from the one given in Ref.~\cite{LiTravesset2019}. This is because $\xi^{\mu}=\bar{\xi}^{\mu}$ so that $g_{\mu \nu} \xi^{\nu}=\xi_{\mu}\ne \bar{\xi}_{\mu}=\bar{g}_{\mu \nu} \xi^{\nu}$, which was overlooked in previous work.

As an example, if the problem has rotational symmetry and the boundary is a circle $r=r_0$, then $\theta(s)=s/r_0$.  The tangent vectors then are 
\begin{equation}
    t^{r}=0 \quad t^{\theta}=\frac{1}{r_0},
\end{equation}
and the normal vectors
\begin{equation}
    n_r = \sqrt{1+f^{\prime}(r_0)^2} \quad n_{\theta}=0.
\end{equation}
Then Eq.~\ref{Eq:tangentvector} becomes
\begin{eqnarray}
    t^{\rho}\nabla_{\rho} t^{r}&=&\Gamma^{r}_{\theta\theta}\frac{1}{r^2_0}=-\frac{1}{r_0(1+f^{\prime}(r_0)^2)} \\\nonumber
    &=& -\frac{1}{r_0\sqrt{1+f^{\prime}(r_0)^2}} n^{r}
\end{eqnarray}
with 
\begin{equation}
    \frac{r_A}{r_0}=\sqrt{1+f^{\prime}(r_0)^2}.
\end{equation}
Finally, the boundary condition becomes equal to,
\begin{equation}\label{Eq:app:boundary condition with rotational symmetry}
    \overline{g}_{rr}(r_0)\sigma^{rr}(r_0)=-\frac{1}{\sqrt{1+f^{\prime}(r_0)^2}}\frac{\tau}{r_0} \ .
\end{equation}
For a sphere $f^{\prime}(r)=-\frac{r}{\sqrt{R^2_0-r^2}}$ and thus
\begin{equation}
    r_A = \frac{r_0}{\sqrt{1-\left(\frac{r_0}{R_0}\right)^2}} \ .
\end{equation}
\end{appendix}

~\\

%

\end{document}


\title{Supplementary information:\\Exact Solution for Elastic Networks on Curved Surfaces}

\author{Yinan Dong}
\affiliation{Department of Physics and Astronomy, University of California Riverside, Riverside, California 92521, United States}

\author{Roya Zandi}
\affiliation{Department of Physics and Astronomy, University of California Riverside, Riverside, California 92521, United States}

\author{Alex~Travesset}
\affiliation{Department of Physics and Astronomy, Iowa State University and Ames Lab, Ames, Iowa 50011, United States}

\pacs{}
\maketitle

\clearpage
\onecolumngrid
\setcounter{figure}{0}
\setcounter{equation}{0}
\setcounter{table}{0}
\renewcommand{\thefigure}{S\arabic{figure}}
\renewcommand{\theequation}{S\arabic{equation}}
\renewcommand{\thetable}{S\arabic{table}}

\section{Free energy Normalization}

We will consider a dimensionless free energy normalized per particle, that is
\begin{equation}\label{Eq:SI:normalization:free}
    f\equiv \frac{F}{Y \bar{A}}=\frac{2F}{\sqrt{3} N Y a^2_L} \ ,
\end{equation}
hence the area in {\em reference space}, see Eq.~3. This area is given by
\begin{equation}
\hat{A}=\pi\left(1-\frac{q_i}{6}\right) \rho_0^2 \ .
\end{equation}
Given two systems with the same number of particles, the one with the smallest free energy per particle, Eq.~\ref{Eq:SI:normalization:free} is the stable minimum.

\section{About units}

The free energy, see Eq.~1 is
\begin{equation}\label{Eq:units:tot}
     F = \int d^2{\bm x}\sqrt{g}\left[{\cal F}^{elastic} + {\cal F}^{bending}\right]+F^{abs}+F^{line} \ .
\end{equation}

Note that the stress tensor, Eq.~A6 is given by
\begin{equation}
    \sigma^{\alpha \beta}= \frac{1}{\sqrt{g}}\frac{\delta F}{\delta u_{\alpha \beta}}=A^{\alpha\beta\gamma\delta}  u_{\gamma \delta} = Y \times \left(\mbox{Terms that Depend on } \nu_p\right)
\end{equation}
Note also, that the ratio of the Young modulus and the line tension defines a coefficient $l_A$ with units of length 
\begin{equation}
    \frac{\tau}{Y}\equiv l_A
\end{equation}
Therefore, through the boundary conditions Eq.~8, the quantity
\begin{equation}
    \frac{\sigma^{\alpha\beta}}{Y}=h(\nu_p,l_A) \ ,
\end{equation}
does not directly depend on the Young modulus. 

Also,
\begin{equation}
    F^{abs}=-N \Delta F = - \frac{2\Delta F}{\sqrt{3}a^2_L} \frac{\sqrt{3}}{2} N a^2_L =-\Pi \hat{A} 
    \mbox{ with } \hat{A}=\int d^2{\bm r}\sqrt{\bar{g}({\bm x})} \ .
\end{equation}
Therefore $\Pi=\frac{2\Delta F}{\sqrt{3}a^2_L}$, and $\hat{A}$ is the area in reference space.

The line tension term is a function of the perimeter ($P$), given by
\begin{equation}
    \frac{F^{line}}{YL^2}=\frac{\tau}{Y L^2} \oint_{\partial D} ds \equiv \frac{\tau P}{Y L^2} \ .
\end{equation}

Finally, the free energy Eq.~\ref{Eq:units:tot} is
\begin{equation}
    \frac{F}{L^2Y}=f(\nu_p,\frac{l_A}{L})+\frac{\kappa}{YL^2}\int d^2{\bm x}\sqrt{g}\left[ \left(\frac{1}{R_1}-H_0\right)^2+ \left(\frac{1}{R_2}-H_0\right)^2\right]+\frac{\Pi}{Y}\frac{\hat{A}}{L^2} + \frac{\tau P}{Y L^2} \ ,
\end{equation}
where $L$ is a characteristic dimension of the system. 
In general we will choose $L^2=\hat{A}$, so
\begin{equation}
     \frac{F}{Y\hat{A}}=f(\nu_p,\frac{l_A}{\sqrt{\hat{A}}})+\frac{\kappa}{Y\hat{A}}\int d^2{\bm x}\sqrt{g}\left[ \left(\frac{1}{R_1}-H_0\right)^2+ \left(\frac{1}{R_2}-H_0\right)^2\right]+\frac{\Pi}{Y} + \frac{\tau P}{Y \hat{A}} \ ,
\end{equation}
which defines the effective linear tension $\hat{\tau}=\frac{\tau a_L}{Y \hat{A}}$ and dimensionless area $\hat{A}/a_L^2$, so that all lengths are expressed in terms of the lattice constant $a_L$.

\section{\label{SI:linear}Connection with Linear Elasticity Theory}

Here we show that the covariant formalism defined by Eq.~1, Eq.~2 reduce to the known formulas from elasticity theory when the displacements are small. Within elasticity theory, the reference metric (without disclinations)
\begin{equation}\label{Eq:SI:LE:reference}
    {\bar g}_{\alpha\beta} = \delta_{\alpha \beta}
\end{equation}
The surface is described in the Monge gauge,
\begin{equation}
    z = h(x,y) \ .
\end{equation}
The mapping ${\bm x}={\cal U}({\bar x})$ is given by
\begin{equation}\label{Eq:SI:LE:mapping}
    {\bm x}=\bar{{\bm x}}+{\bm u}(\bar{{\bm x}}) \ ,
\end{equation}
where ${\bm u}$ is the displacement. Then, the actual metric becomes
\begin{eqnarray}\label{Eq:SI:LE:actual}
    g_{\alpha\beta}(\bar{\bm x}) &=& \bar{\partial}_{\alpha} {\vec r}({\bar {\bm x}}) \bar{\partial}_{\beta} {\vec r}({\bar {\bm x}})=\delta_{\alpha\beta}+\bar{\partial}_{\alpha} u_{\beta}+\bar{\partial}_{\beta} u_{\alpha}+ \bar{\partial}_{\alpha} u_{\gamma} \bar{\partial}_{\beta} u_{\gamma}+{\bar \partial}_{\rho} h {\bar \partial}_{\gamma} h\left(\delta_{\alpha\rho}\delta_{\gamma\beta}+\delta_{\alpha\rho}\bar{\partial}_{\beta}u_{\gamma}+\bar{\partial}_{\alpha}u_{\rho}\delta_{\beta\lambda}+ \bar{\partial}_{\alpha}u_{\rho}\bar{\partial}_{\beta}u_{\gamma}\right)
    \nonumber\\
     &\approx& \delta_{\alpha\beta}+\bar{\partial}_{\alpha} u_{\beta}+\bar{\partial}_{\beta} u_{\alpha}+{\bar \partial}_{\alpha} h {\bar \partial}_{\beta} h
\end{eqnarray}
If only linear terms in ${\bm u}$ and the leading term in $h$ are kept the
strain tensor Eq.~2 becomes
\begin{equation}\label{Eq:SI:LE:strain}
    u_{\alpha \beta}=\frac{1}{2}\left(\bar{\partial}_{\alpha} u_{\beta}+\bar{\partial}_{\beta} u_{\alpha}+{\bar \partial}_{\alpha} h {\bar \partial}_{\beta} h \right) \ .
\end{equation}
The actual metric is
\begin{equation}
    g_{\alpha\beta}({\bar{\bm x}})=\bar{g}_{\alpha\beta}({\bar{\bm x}})+2 u_{\alpha \beta}({\bar{\bm x}})
\end{equation}
The leading elastic part of the free energy, consistent with the expansion Eq.~\ref{Eq:SI:LE:strain} becomes
\begin{eqnarray}\label{Eq:SI:LE:free_energy}
{\cal F}^{elastic}&=&\frac{1}{2}\frac{Y}{1-\nu_p^2}\left(\nu_p(u_{\alpha\alpha}^2+(1-\nu_p)u_{\alpha\beta} u_{\alpha\beta} \right)
\nonumber \\
&=& \frac{1}{2} \left( 2\mu (u_{\alpha\beta})^2+\lambda (u_{\alpha\alpha})^2\right) 
\end{eqnarray}
expressed in terms of the Lame coefficients $\lambda,\mu$ instead of the Young modulus $Y=\frac{4\mu(\mu+\lambda)}{2\mu+\lambda}$ and Poisson ratio $\nu_p=\frac{\lambda}{2\mu+\lambda}$. For fixed geometry, that is for a given $f$, this is exactly the same free energy and strains as used in linear elasticity theory, see for example Ref.~\cite{Seung1988}. The Airy function is the solution to the equation, 
\begin{equation}\label{Eq:SI:LE:Airy}
    \frac{1}{Y_0}{\bar{\Delta}}^2 \chi(\bar{{\bm x}})=s({\bar{\bm x}})-K({\bar{\bm x}}) \ ,
\end{equation}
see Ref.~\cite{Seung1988} for a full derivation. The laplacian ${\bar{\Delta}}$ refers to a flat metric. Adding an arbitrary disclination density is done by introducing singularities in the reference metric, as discussed for a central disclination in the main text. It is
\begin{equation}\label{Eq:SI:LE:disc_density}
s({\bar{\bm x}})=\frac{\pi}{3}\sum_{i=1}^N q_i \delta({\bar{\bm x}}-{\bar{\bm x}}_i)
\end{equation}
where ${\bar x}_i$ are the positions of the $N$ disclinations. The Gaussian curvature is obtained by expanding Eq.~A3 to leading order, consistent with the expansion in the actual metric Eq.~\ref{Eq:SI:LE:actual}. Therefore
\begin{equation}\label{Eq:SI:LE:Gaussian_curvature}
    K({\bar{\bm x}}) = -\frac{1}{2}\varepsilon_{\alpha \beta}\varepsilon_{\gamma \rho} \bar{\partial}_{\beta}  \bar{\partial}_{\rho} \left(\bar{\partial}_{\alpha} h \bar{\partial}_{\gamma} h\right) \ .
\end{equation}

We remark that Eq.~\ref{Eq:SI:LE:Airy} is written in terms of a flat metric, and the only contribution from the curved surface is through the approximated Gaussian curvature Eq.~\ref{Eq:SI:LE:Gaussian_curvature}. Eq.~\ref{Eq:SI:LE:Airy} has the physical interpretation of the Gaussian curvature screening the disclination density.

The explicit form of the stress tensor is obtained from the Airy function as 
\begin{eqnarray}\label{Eq:SI:LE:stress}
 \sigma^{\rho\rho}(\rho)&=&\frac{1}{\rho}\frac{d\chi(\rho)}{d\rho}
 \\\nonumber
 &=& \frac{Y}{2\rho^2}\left(\frac{q_i}{6}\rho^2\log\left(\frac{\rho}{\rho_0}\right)+ \int^{\rho}_{0} dv v \int_{v}^{\rho_0} \frac{du}{u}\left(\frac{df}{du}\right)^2-\frac{\rho^2}{\rho_0^2}\int^{\rho_0}_{0} dv v \int_{v}^{\rho_0} \frac{du}{u}\left(\frac{df}{du}\right)^2\right) \\\nonumber
    \sigma^{\theta \theta}(\rho)&=&\frac{1}{\rho^2}\frac{d^2\chi(\rho)}{d\rho^2}=\frac{1}{\rho^2}\frac{d(\rho \sigma^{\rho\rho}(\rho))}{d\rho}  \\ \nonumber
    &=& \frac{Y}{2 \rho^2}\left( \frac{q_i}{6}\log\left(e\frac{\rho}{\rho_0}\right)+\int_{\rho}^{\rho_0} \frac{du}{u}\left(\frac{df}{du}\right)^2-\frac{1}{\rho^2}\int^{\rho}_{0} dv v \int_{v}^{\rho_0} \frac{du}{u}\left(\frac{df}{du}\right)^2 - \frac{1}{\rho_0^2}\int^{\rho_0}_{0} dv v \int_{v}^{\rho_0} \frac{du}{u}\left(\frac{df}{du}\right)^2\right) \ .
\end{eqnarray}
The strain tensor is
\begin{eqnarray}\label{Eq:SI:LE:strain_explicit}
    u_{\rho\rho}&=&\frac{1}{Y}\left(\sigma^{\rho\rho}-\nu_p\rho^2\sigma^{\theta\theta}\right)=\frac{1}{Y}\left(\frac{1}{\rho}\frac{d}{d\rho} -\nu_p\frac{d^2}{d^2\rho}\right)\chi \nonumber\\
    u_{\theta\theta}&=&\frac{\rho^2}{Y}\left(\rho^2\sigma^{\theta\theta}-\nu_p\sigma^{\rho\rho}\right)=\frac{\rho^2}{Y}\left( \frac{d^2}{d^2\rho}-\frac{\nu_p}{\rho}\frac{d}{d\rho}\right)\chi  \ .
\end{eqnarray}
Note that these equations are also equivalent to Eq.~\ref{Eq:SI:LE:strain}
\begin{eqnarray}\label{Eq:SI:LE:strain_cylindrical}
    u_{\rho\rho}&=&\frac{d u_\rho}{d\rho}+\frac{1}{2}\partial_{\rho}h \partial_{\rho}h \\\nonumber
    \frac{u_{\theta\theta}}{\rho^2}&=& -  {\Gamma}^{\rho}_{\theta \theta} u_{\rho}=\frac{u_{\rho}}{\rho}
\end{eqnarray}
with $r(\rho)=\rho+u_{\rho}(\rho)$. The free energy is
\begin{equation}\label{Eq:SI:LE:free_energy_explicit}
    F=\pi\int_0^{\rho_0} d\rho \rho \left( \sigma^{\rho\rho}u_{\rho\rho}+\sigma^{\theta\theta}u_{\theta\theta}\right) \ .
\end{equation}
Finally, the mapping $r(\rho)$ (or $\rho(r))$ can be obtained from solving either equation
\begin{eqnarray}\label{Eq:SI:LE:mapping_equation}
    2u_{\rho\rho}(\rho)&=&(1+f^{\prime}(r)^2)\left(\frac{dr}{d\rho}\right)^2-1 \\\nonumber
    2u_{\theta\theta}(\rho)&=&r^2(\rho)-\alpha^2\rho^2 \ .
\end{eqnarray}
The explicit solution of the previous equation, consistent at linear order is
\begin{equation}\label{Eq:SI:LE:linear_solution}
    r(\rho)=\rho-(1-\alpha)\rho+\frac{u_{\theta\theta}(\rho)}{ \rho}=\rho\left(1-\frac{q_i}{6}+\frac{1}{Y}\left[ \frac{d^2}{d\rho^2}-\frac{\nu_p}{\rho}\frac{d}{d\rho}\right]\chi
    \right) \ ,
\end{equation}
Note that had we used the first equation, the solution
\begin{eqnarray}
   \int_0^r dr\sqrt{1+f^{\prime}(r)^2}dr &=& \int_0^{\rho} d\rho\sqrt{1+2u_{\rho\rho}} \nonumber \\
   r+\frac{1}{2}\int_0^r f^{\prime}(r)^2 &=& \rho+\int_0^{\rho} d\rho u_{\rho\rho} \nonumber \\
   r&=& \rho +\int_0^{\rho} d\rho\frac{d u_{\rho}}{d\rho}=\rho+u_{\rho} \ ,
\end{eqnarray}
where Eq.~\ref{Eq:SI:LE:strain_cylindrical} has been used.

Addition of a line tension just adds the boundary condition
\begin{equation}\label{Eq:SI:LE:sigma_line}
    \sigma^{\rho\rho}=-\frac{\tau}{\rho_0} 
\end{equation}
to the stress tensor, and the additional free energy contribution
\begin{equation}\label{Eq:SI:LE:free_line}
    F^{line} = 2\pi \tau r(\rho_0) \equiv 2\pi \tau r_0 \ .
\end{equation}

Analytical formulas for the Airy function, stress tensor, strain tensor, and free energy for two different surfaces, the spheroid and sombrero are given below.

\subsubsection{Spheroid surface}

It is given by (with $R_0$ the spheroid radius, not to be confused with $r_0=\rho(r_0)$ the coordinate parameterizing the boundary) the equation 
\begin{equation}
    f(r)=\beta \sqrt{R_0^2-r^2} \ .
\end{equation}
The Airy function is 

\begin{eqnarray}
    \chi&=&Y\frac{\beta^2}{16\rho_0^2}\bigg(-R_0^2\log\big(R_0^2\big) \bigg(\rho^2-2\rho_0^2\log\big(\rho\big)\bigg)+\rho_0^2 \left(-\rho^2-\left(R_0^2-\rho^2+R_0^2\log\bigg(\frac{\rho^2}{R_0^2}\bigg)\right)\log\big(R_0^2-\rho^2\big)\right)+\nonumber\\
    &+&\rho^2\big(R_0^2-\rho_0^2\big)\log\big(R_0^2-\rho_0^2\big)-R_0^2\rho_0^2 Li_{2}\left(1-\frac{\rho^2}{R_0^2}\right)\bigg)- \frac{\tau\rho^2}{2 r_A} \ .
\end{eqnarray}
where $Li_2$ is the polylogarithm function $Li_2(x)=\sum_{n=1}^{\infty} \frac{x^n}{n^2}$.
\\
The stress tensors is 
\begin{eqnarray}
    \sigma^{\rho\rho}&=&Y\frac{1}{24\rho^2\rho_0^2}(3R_0^2\beta^2(\rho_0^2-\rho^2)\log R_0^2+3\beta^2(\rho^2-R_0^2)\rho_0^2\log(R_0^2-\rho^2)+
    \nonumber\\
    &+&\rho^2(2q_i\rho_0^2\log\left(\frac{\rho}{\rho_0}\right)+3\beta^2(R_0^2-\rho_0^2)\log(R_0^2-\rho_0^2)))- \frac{\tau}{r_A }\nonumber \\
    \sigma^{\theta\theta}&=&Y\frac{1}{24\rho^4\rho^2}(-3R_0^2\beta^2(\rho^2+\rho_0^2)\log R_0^2 + 3\beta^2(R_0^2+\rho^2)\rho_0^2\log (R_0^2-\rho^2) \nonumber\\
    &+&\rho^2(2\rho_0^2(q_i+3\beta^2+q_i\log\left(\frac{\rho}{\rho_0}\right))+ 3\beta^2(R_0^2 - \rho_0^2)\log(R_0^2-\rho_0^2)))- \frac{\tau}{r_A }\frac{1}{\rho^2} \ .
\end{eqnarray}
The strain tensors is
\begin{eqnarray}
    u_{\rho\rho}&=&\frac{1}{24\rho^2\rho_0^2}(3R_0^2\beta^2((-1+\nu_p)\rho^2)\log(R_0^2) - 3\beta^2(R_0^2(1+\nu_p)+\nonumber\\
    &+&(-1+\nu_p)\rho^2)\rho_0^2\log(R_0^2-\rho^2) + \rho^2(-2\rho_0^2(3\nu_p\beta^2-q_i\log\left(\frac{\rho}{\rho_0}\right)+q_i\nu_p\log\left(\frac{e\rho}{\rho_0}\right)) +\nonumber\\
    &+& 3(-1+\nu_p)\beta^2(\rho_0^2-R_0^2)\log(R_0^2-\rho_0^2)))+ (\nu_p - 1)\frac{\tau}{Y r_A}\nonumber \\
    u_{\theta\theta}&=&\frac{1}{24\rho_0^2}(3R_0^2\beta^2((-1+\nu_p)\rho^2 - (1+\nu_p)\rho_0^2)\log(R_0^2) + 3\beta^2(R_0^2(1+\nu_p)-\nonumber\\
    &-& (-1+\nu_p)\rho^2)\rho_0^2\log(R_0^2-\rho^2) + \rho^2(2\rho_0^2(q_i+3\beta^2-q_i(-1+\nu_p)\log\left(\frac{\rho}{\rho_0}\right) +\nonumber \\ 
    &+&3(-1+\nu_p)\beta^2(\rho_0^2-R_0^2)\log(R_0^2-\rho_0^2)))+ \rho^2(\nu_p - 1)\frac{\tau}{Y r_A} \ .
\end{eqnarray}

If we expand $f(r)$ in powers of the spheroid radius $R_0$,
\begin{equation}
    f(r)=\beta R_0(1-\frac{r^2}{2R_0^2}) \ .
\end{equation} 
The Airy function is
\begin{eqnarray}
    \chi = -\frac{Y\beta^2 (\rho^4-2\rho^2\rho_0^2)}{64 R_0^2} + \frac{Y q_i}{24}\rho^2\left(\ln\left(\frac{\rho}{\rho_0}\right)-\frac{1}{2}\right) - \frac{\tau\rho^2}{2 r_A} \ .
\end{eqnarray}
The stress tensors is
\begin{eqnarray}
    \sigma^{\rho\rho}=\frac{Y\beta^2}{16R_0^2}(\rho_0^2-\rho^2)+\frac{Y}{2\rho^2}\left(\frac{q_i}{6}\rho^2 \log\left(\frac{\rho}{\rho_0}\right)\right) - \frac{\tau}{r_A} \\
    \sigma^{\theta\theta}=\frac{Y\beta^2}{16 R_0^2 \rho^2}(\rho_0^2-3\rho^2)+\frac{Y}{2\rho^2}\left(\frac{q_i}{6}\log\left(e\frac{\rho}{\rho_0}\right)\right) - \frac{\tau}{r_A }\frac{1}{\rho^2} \ .
\end{eqnarray}
The strain tensors is
\begin{eqnarray}
    u_{\rho\rho}&=&\frac{\beta^2}{16R_0^2}\left((3\nu_p - 1)\rho^2-(\nu_p-1)\rho_0^2\right)+\frac{1}{12}q_i\log\left(\frac{\rho}{\rho_0}\right)-\frac{1}{12}q_i\nu_p\log\left(\frac{e\rho}{\rho_0}\right) + (\nu_p-1)\frac{\tau}{Y r_A}\\\nonumber
    u_{\theta\theta}&=&\frac{\beta^2\rho^2}{16R_0^2}\left((\nu_p-3)\rho^2-(\nu_p-1)\rho_0^2\right)+\frac{\rho^2}{12}\left(-\nu_p q_i\log\left(\frac{\rho}{\rho0}\right)+q_i\log\left(\frac{e\rho}{\rho_0}\right)\right) + \rho^2(\nu_p - 1)\frac{\tau}{Y r_A} \ .
\end{eqnarray}
The free energy is
\begin{eqnarray}
    F=\pi\left( Y\left(\frac{q_i^2\rho_0^2}{288}-\frac{q_i^2\beta^2\rho_0^4}{192R_0^2}+\frac{\beta^4\rho_0^6}{384R_0^4}\right) + \frac{(1-\nu_p)\rho_0^2\tau^2}{Y r_A^2}\right) \ .
\end{eqnarray}

\subsubsection{Sombrero surface}

It is given by the equation
\begin{equation}
f(r)=\frac{\beta R_0}{3}\left(1-\left(\frac{r}{R_0}\right)^2+\left(\frac{r}{R_0}\right)^4\right)^{3/2} \ 
\end{equation}
where $r$ is the radius of the sombrero surface. 
The Airy function is 
\begin{eqnarray}
    \chi&=&-\frac{Y\beta^2\rho^2}{57600 {R_0}^{10}}\bigg(900 {R_0}^8(\rho^2 - 2{\rho_0}^2) - 1000 {R_0}^6(\rho^4 - 3{\rho_0}^4) + 675 {R_0}^4(\rho^6-4{\rho_0}^6) - \nonumber \\
    &-&288 {R_0}^2(\rho^8-5{\rho_0}^8) + 80 (\rho^{10}-6{\rho_0}^{10})\bigg)  - \frac{\tau\rho^2}{2 r_A} \
\end{eqnarray}
The stress tensor is
\begin{eqnarray}
     \sigma^{\rho\rho}&=&\frac{Y\beta^2}{480}\bigg(\frac{30(-\rho^2+\rho_0^2)}{R_0^2} + \frac{50(\rho^4 - \rho_0^4)}{R_0^4} + \frac{45(-\rho^6 + \rho_0^6)}{R_0^6} + \frac{24(\rho^8 - \rho_0^8)}{R_0^8} + \nonumber \\
     &+& \frac{8(-\rho^{10} + \rho_0^{10})}{R_0^{10}}\bigg) 
     +\frac{Yq_i}{12}\log\bigg(\frac{\rho}{\rho_0}\bigg) - \frac{\tau}{r_A}\\
     \sigma^{\theta\theta}&=&\frac{Y\beta^2}{480\rho^2}\bigg(\frac{30(-3\rho^2+\rho_0^2)}{R_0^2}+\frac{50(5\rho^4-\rho_0^4)}{R_0^4}-\frac{45(7\rho^6-\rho_0^6)}{R_0^6}+\frac{24(9\rho^8-\rho_0^8)}{R_0^8}+ \nonumber \\ &+&\frac{8(-11\rho^{10}+\rho_0^{10})}{R_0^{10}}\bigg) + \frac{Yq_i}{12\rho^2}\log\bigg(\frac{e\rho}{\rho_0}\bigg) - \frac{\tau}{r_A\rho^2}
\end{eqnarray}
The strain tensor is
\begin{eqnarray}
    u_{\rho\rho}&=&\frac{\beta^2}{480}\bigg(\frac{30((-1+3\nu_p)\rho^2-(-1+\nu_p)\rho_0^2)}{R_0^2} - 50\frac{(-1+5\nu_p)\rho^4-(-1+\nu_p)\rho_0^4}{R_0^4}+\nonumber\\
    &+&\frac{45(-1+7\nu_p)\rho^6-(-1+\nu_p)\rho_0^6}{R_0^6} - \frac{24(-1+9\nu_p)\rho^8-(-1+\nu_p\rho_0^8)}{R_0^8}+\nonumber\\
    &+&\frac{8((-1+11\nu_p)\rho^{10}-(-1+\nu_p)\rho_0^{10})}{R_0^{10}}\bigg) + \frac{q_i}{12}\bigg(\log\bigg(\frac{\rho}{\rho_0}\bigg) - \nu_p\log\bigg(\frac{e\rho}{\rho_0}\bigg)\bigg)+\frac{(-1+\nu_p)\tau}{Yr_A}\\
    u_{\theta\theta}&=&\frac{\rho^2\beta^2}{480}\bigg(\frac{30((-3+\nu_p)\rho^2-(-1+\nu_p)\rho_0^2)}{R_0^2} - \frac{50((-5+\nu_p)\rho^4 - (-1+\nu_p)\rho_0^4)}{R_0^4}  + \nonumber\\
    &+&\frac{45((-7+\nu_p)\rho^{6}-(-1+\nu_p)\rho_0^{6})}{R_0^6}-\frac{24((-9+\nu_p)\rho^8-(-1+\nu_p)\rho_0^8)}{R_0^8}+\nonumber\\
    &+&\frac{8((-11+\nu_p)\rho^{10})-(-1+\nu_p))\rho_0^{10}}{R_0^{10}}\bigg) + \frac{q_i}{12}\bigg(\log\bigg(\frac{e\rho}{\rho_0}\bigg)-\nu_p log\bigg(\frac{\rho}{\rho_0}\bigg)\bigg)\bigg)+\frac{(-1 + \nu_p)\tau\rho^2}{Yr_A}
\end{eqnarray}
The free energy is 
\begin{eqnarray}
    F&=&\pi\bigg(Y\bigg(\frac{q_i^2\rho_0^2}{288} - \frac{q_i\beta^2\rho_0^4(900R_0^8-2000R_0^6\rho_0^2+2025R_0^4\rho_0^4-1152R_0^2\rho_0^6+400R_0^8)}{172800R_0^{10}} + \nonumber \\ &+&\frac{\beta^4\rho_0^6}{8870400R_0^{20}}(23100R_0^{16}-115500R_0{14}\rho_0^{14}\rho_0^2+278740R_0^{12}\rho_0^4-420420R_0^{10}\rho_0^6+438075R_0^8\rho_0^8- \nonumber\\
    &-&326480R_0^6\rho_0{10}+171248R_0^4\rho_0^{12}-59136R_0^2\rho_0^{14}+11200\rho_0^{16})\bigg)-\frac{(-1+\nu_p)\rho_0^2\tau^2}{Yr_A^2}\bigg)
\end{eqnarray}

\section{Theory of Defects, inverse Laplacian square}

We finally note that Eq.~\ref{Eq:SI:LE:Airy} can be promoted to an equation 
\begin{equation}\label{Eq:SI:LE:effective_defects}
    \Delta^2 \chi({\bm x}) = s({\bm x})-K({\bm x})
\end{equation}
in terms of the actual metric. In this case, the values used for the Gaussian curvature and the disclination density are covariant and exact. This is the starting point of the effective theory of defects. For rotational symmetric cases, it is possible to solve the equation exactly, at least by numerical integration.

\section{\label{SI:plots}Additional Plots}

\begin{figure}[ht]
  \includegraphics[width=.8\linewidth]{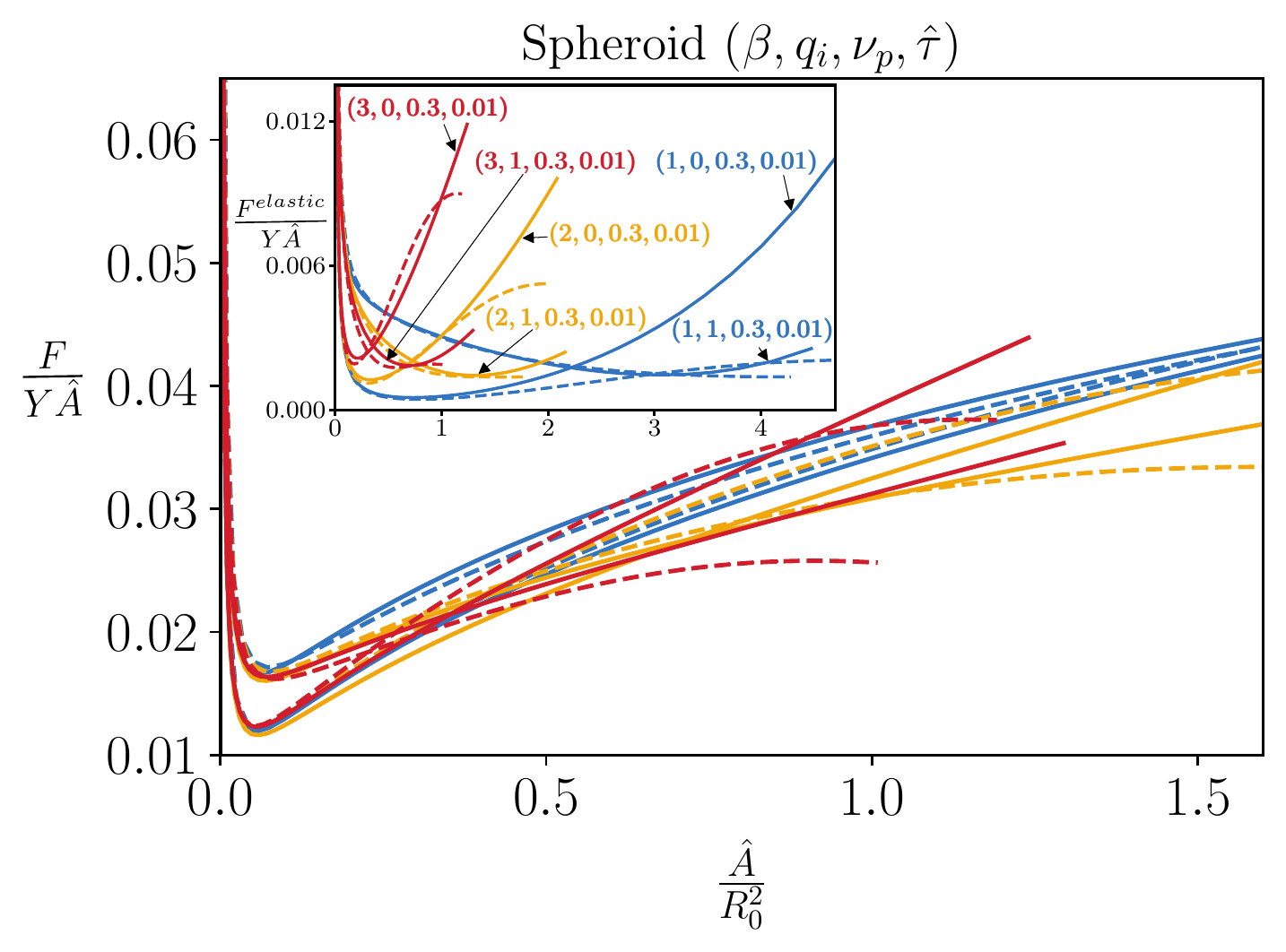} 
  \includegraphics[width=.8\linewidth]{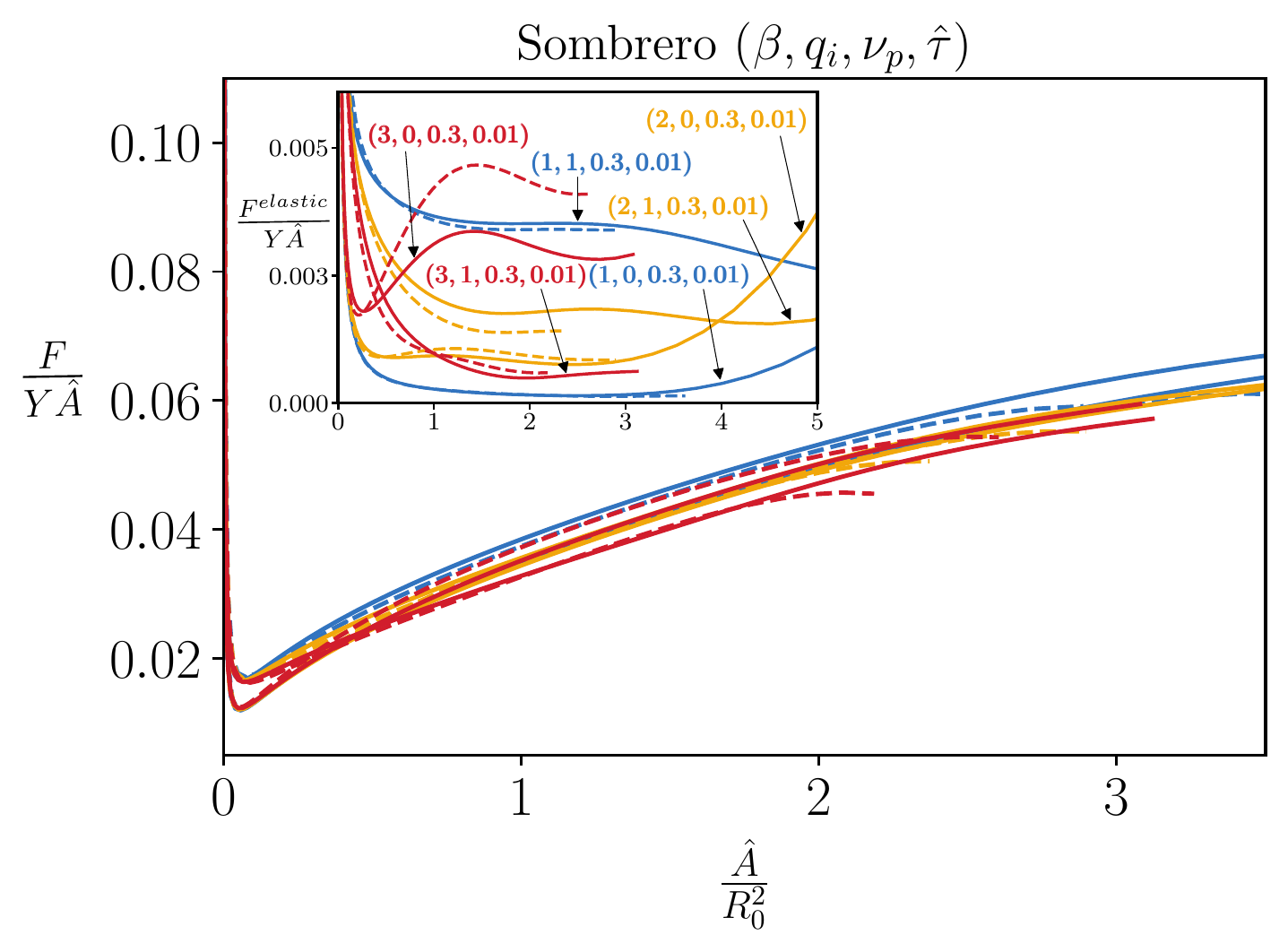}
\caption{Free energy Eq.~1 for a spheroid and sombrero at $\hat{\tau}=0.01$. The solid lines correspond to the exact results, while the dashed lines denote analytical results within linear elasticity.}
\label{fig:SI_tau00p1}
\end{figure}

\begin{figure}[ht]
  \includegraphics[width=.8\linewidth]{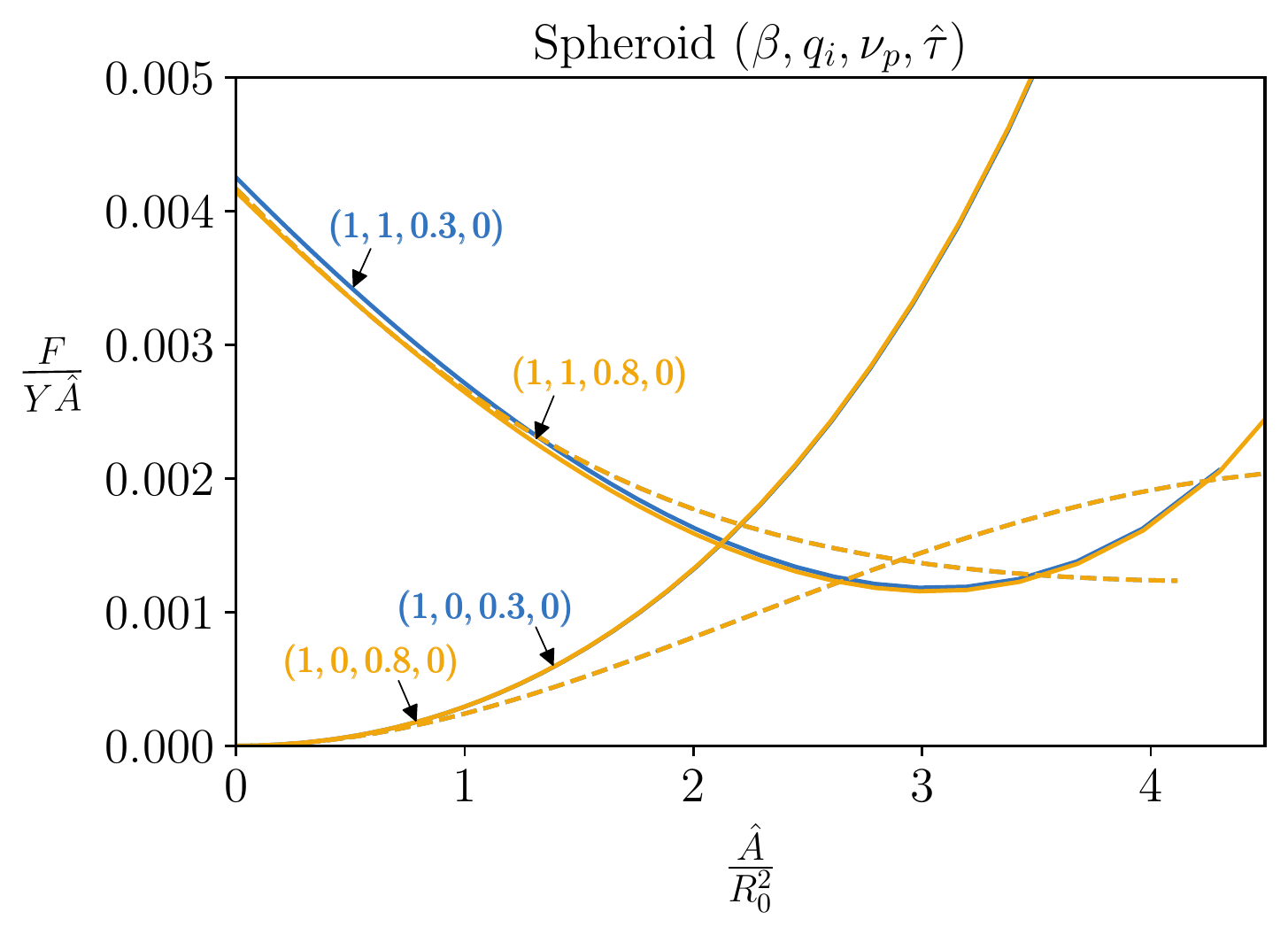} 
  \includegraphics[width=.8\linewidth]{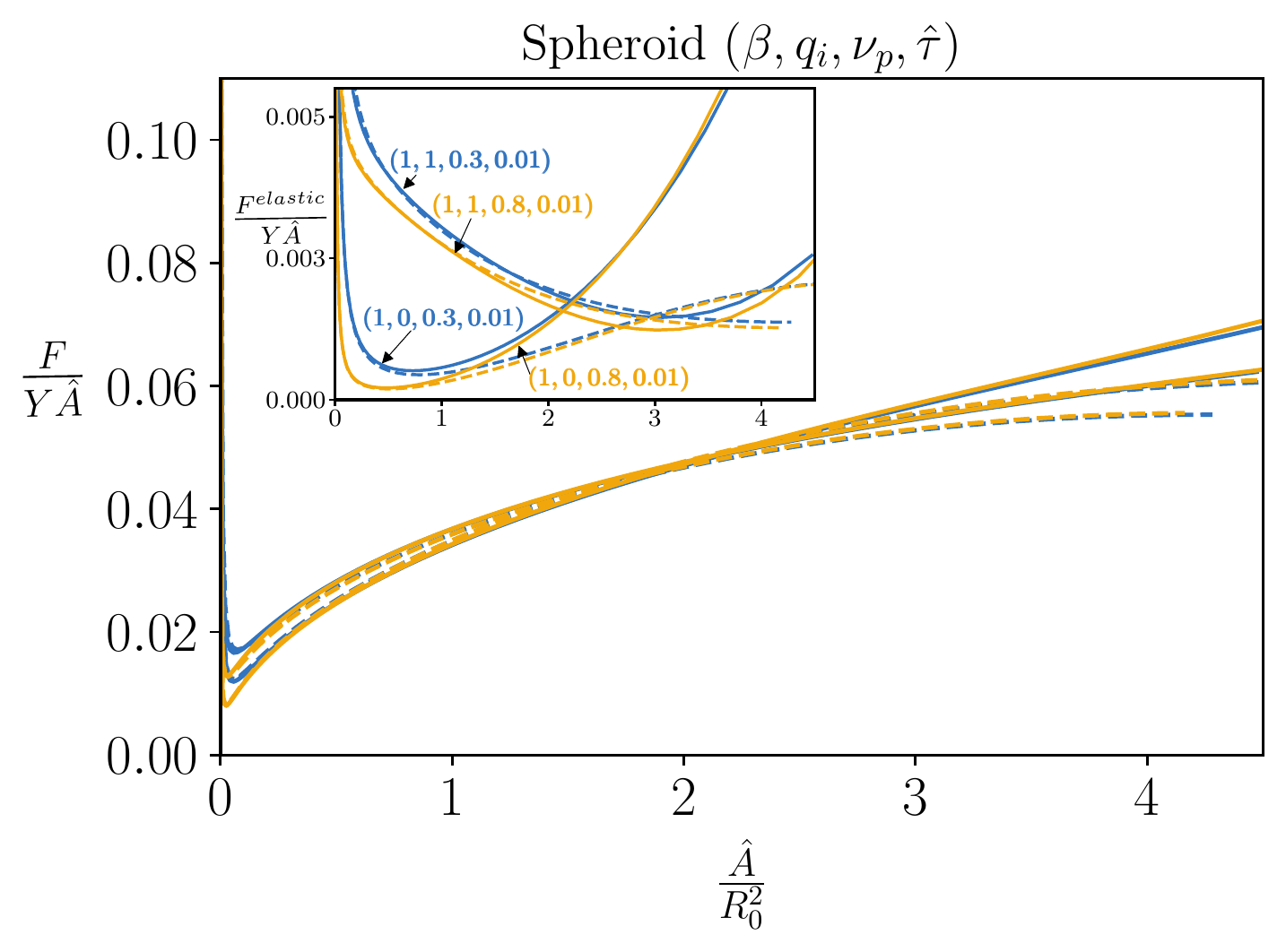}
\caption{Free energy Eq.~1 for the spheroid at different Poisson ratios $\nu_p = 0.3 and 0.8$. The solid lines correspond to the exact results while the dashed lines denote analytical results within linear elasticity.}
\label{fig:SI_tau_different_poissonratio}
\end{figure}

\begin{figure}[ht]
  \includegraphics[width=.8\linewidth]{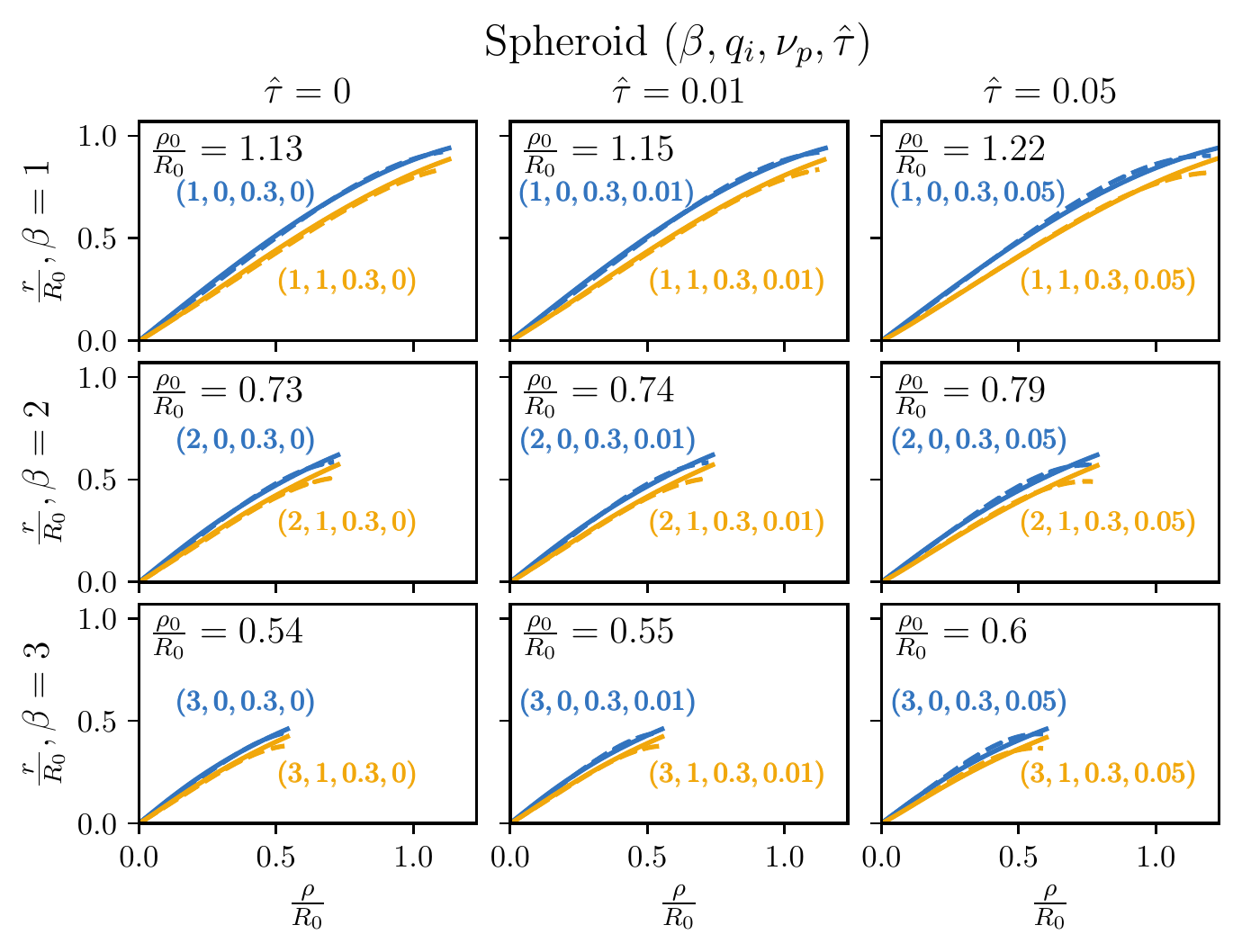}
  \includegraphics[width=.8\linewidth]{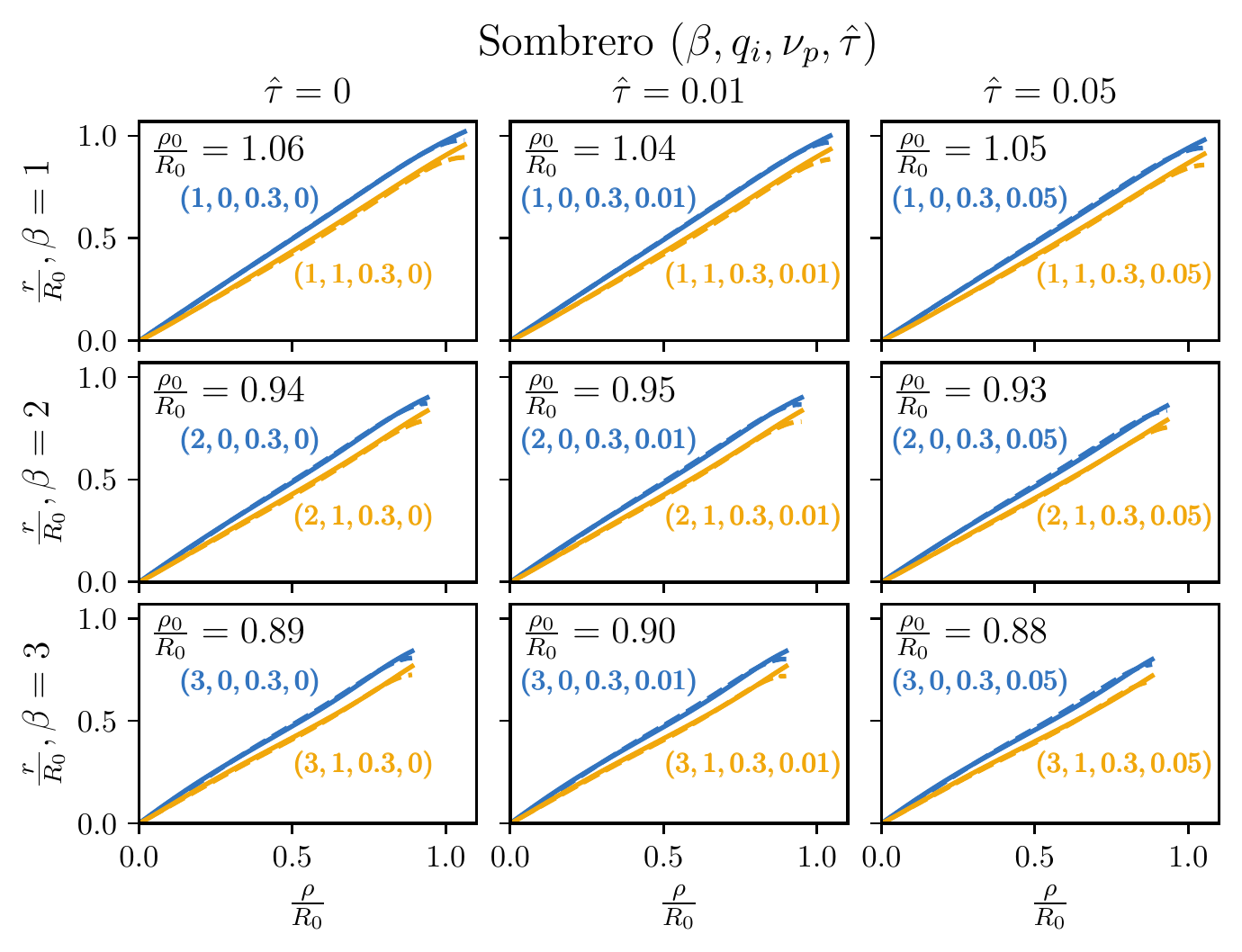}
\caption{Solution Eq.~B4 and Eq.~\ref{Eq:SI:LE:linear_solution} for the spheroid and sombrero relatively small $\rho_0$ compared with FIG.~\ref{fig:SI_solution_big_rho}. The solid lines correspond to the exact result while the dashed lines denote analytical results within linear elasticity.}
\label{fig:SI_solution_rho}
\end{figure}

\begin{figure}[ht]
  \includegraphics[width=.8\linewidth]{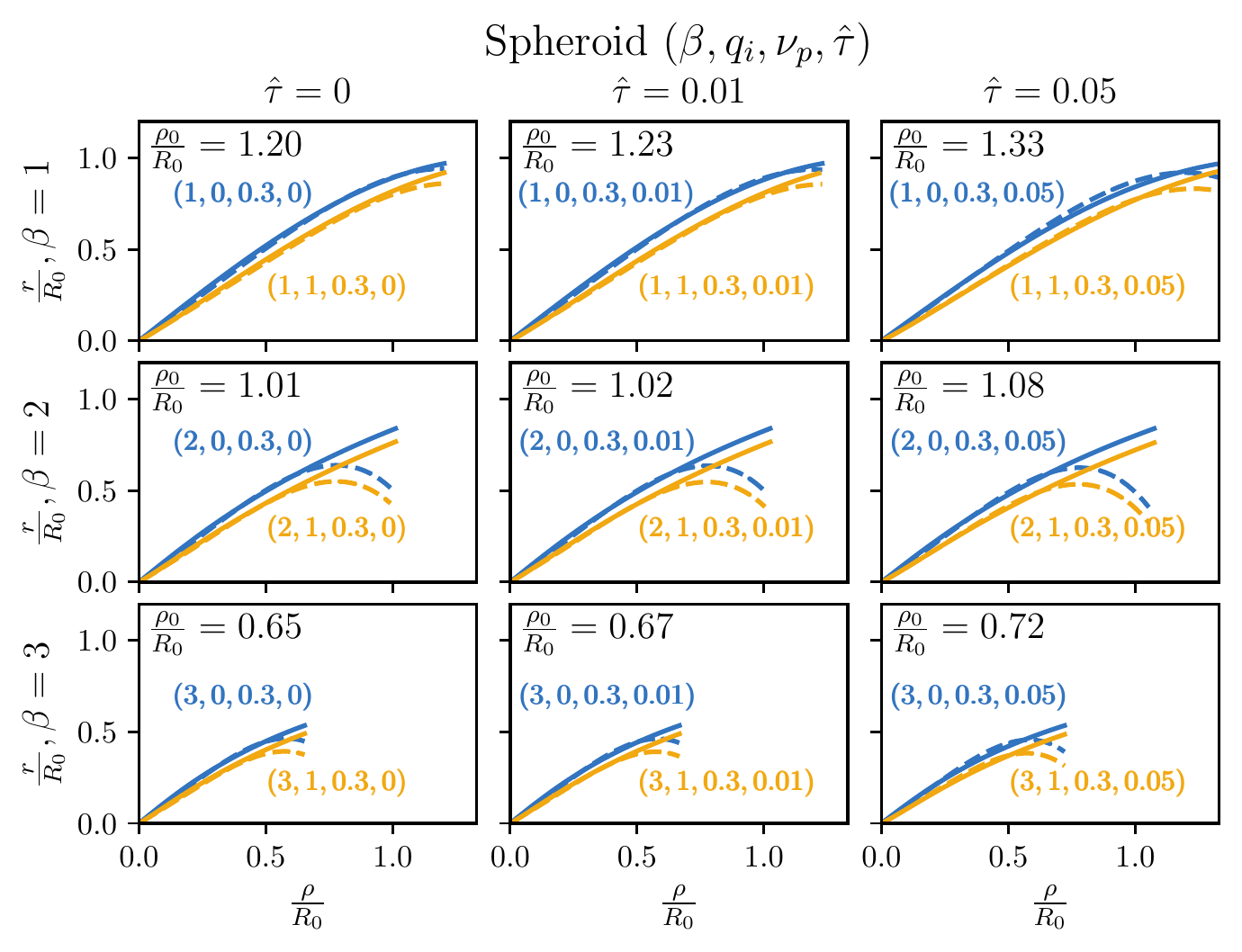}
  \includegraphics[width=.8\linewidth]{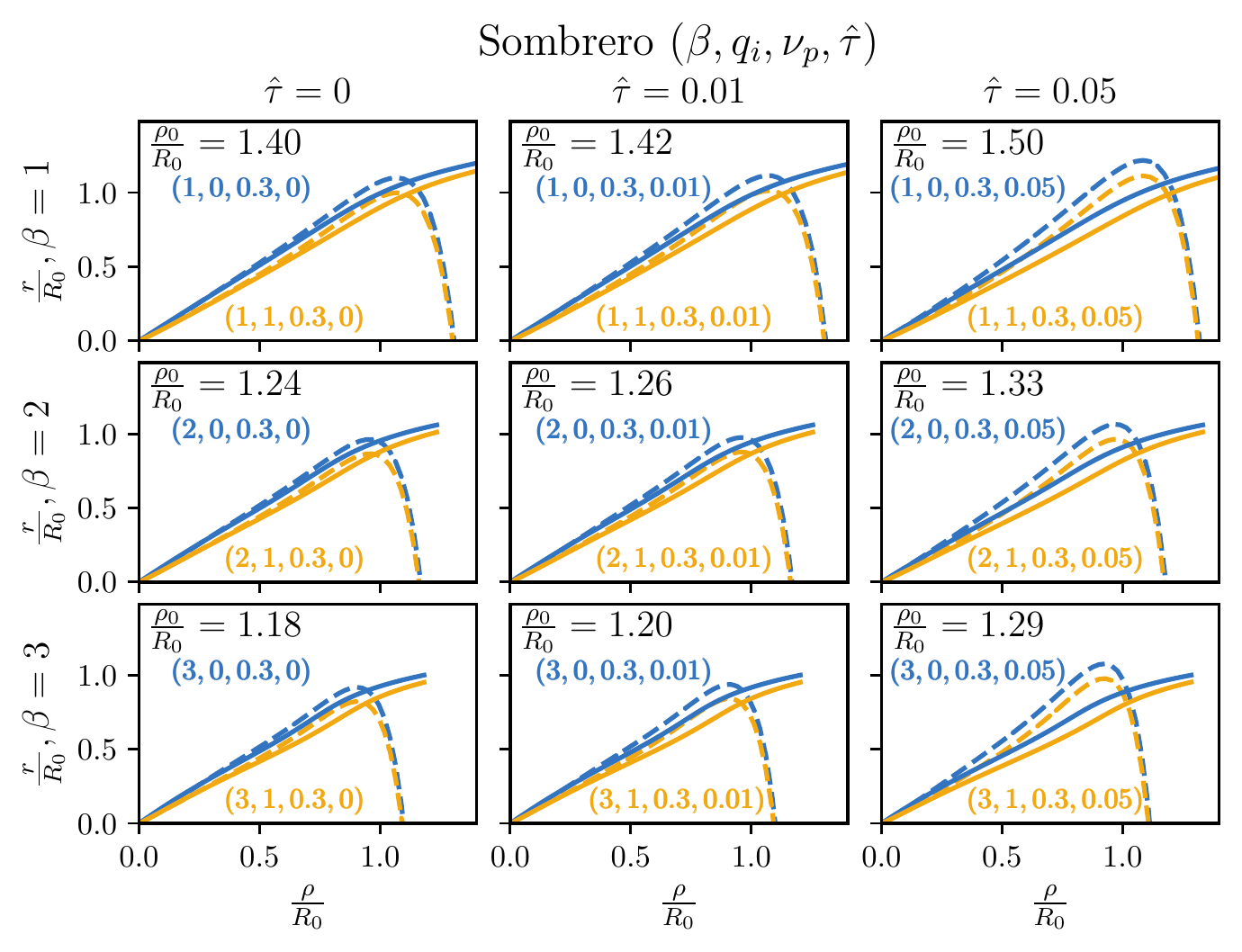}
\caption{Solution Eq.~B4 and Eq.~\ref{Eq:SI:LE:linear_solution} for the spheroid and sombrero with relatively large $\rho_0$ compared with FIG.~\ref{fig:SI_solution_rho}. The solid lines correspond to the exact results while the dashed lines denote analytical results within linear elasticity.}
\label{fig:SI_solution_big_rho}
\end{figure}

\clearpage
%